\documentclass[twocolumn]{aastex701}

\usepackage{tabularx} 
\usepackage{booktabs}
\usepackage{threeparttable} 
\usepackage{multirow} 
\usepackage{graphicx}
\usepackage{amsmath} 
\usepackage{amssymb}
\usepackage{subfigure}
\bibpunct{(}{)}{;}{a}{}{,}  
\makeatletter
\renewcommand{\@thesubfigure}{\hskip\subfiglabelskip}
\makeatother
\makeatletter

\graphicspath{{./}{Fig/}}

\shorttitle{Validating the ICCF-Cut Method for H$\alpha$ Reverberation Mapping}
\shortauthors{Wang et al.}

\begin{document}
\title{Validating the ICCF-Cut Method with Simultaneous Photometric and Spectroscopic H$\alpha$ Reverberation Mapping of NGC 4151 and UGC 3374}
\author[0009-0002-5955-4932]{Chi-Zhuo Wang}
\affiliation{Department of Astronomy, School of Physics, Peking University, Beijing 100871, People's Republic of China}
\affiliation{Kavli Institute for Astronomy and Astrophysics, Peking University, Beijing 100871, People's Republic of China}
\email{chizhuowang@stu.pku.edu.cn}

\author[0000-0002-1530-2680]{Hai-Cheng Feng}
\affiliation{Yunnan Observatories, Chinese Academy of Sciences, Kunming 650216, Yunnan, People’s Republic of China}
\affiliation{Key Laboratory for the Structure and Evolution of Celestial Objects, Chinese Academy of Sciences, Kunming 650216, Yunnan, People’s Republic of China}
\affiliation{Center for Astronomical Mega-Science, Chinese Academy of Sciences, 20A Datun Road, Chaoyang District, Beijing 100012, People’s Republic of China}
\affiliation{Key Laboratory of Radio Astronomy and Technology, Chinese Academy of Sciences, 20A Datun Road, Chaoyang District, Beijing 100101, People’s Republic of China}
\email{hcfeng@ynao.ac.cn}

\author[0000-0003-3823-3419]{Sha-Sha Li}
\affiliation{Yunnan Observatories, Chinese Academy of Sciences, Kunming 650216, Yunnan, People’s Republic of China}
\affiliation{Key Laboratory for the Structure and Evolution of Celestial Objects, Chinese Academy of Sciences, Kunming 650216, Yunnan, People’s Republic of China}
\affiliation{Center for Astronomical Mega-Science, Chinese Academy of Sciences, 20A Datun Road, Chaoyang District, Beijing 100012, People’s Republic of China}
\affiliation{Key Laboratory of Radio Astronomy and Technology, Chinese Academy of Sciences, 20A Datun Road, Chaoyang District, Beijing 100101, People’s Republic of China}
\email{lishasha@ynao.ac.cn}

\author[0000-0002-7350-6913]{Xue-Bing Wu}
\affiliation{Department of Astronomy, School of Physics, Peking University, Beijing 100871, People's Republic of China}
\affiliation{Kavli Institute for Astronomy and Astrophysics, Peking University, Beijing 100871, People's Republic of China}
\email{wuxb@pku.edu.cn}

\author[0000-0003-0827-2273]{Qinchun Ma}
\affiliation{Department of Astronomy, School of Physics, Peking University, Beijing 100871, People's Republic of China}
\affiliation{Kavli Institute for Astronomy and Astrophysics, Peking University, Beijing 100871, People's Republic of China}
\email{maqinchun@pku.edu.cn}


\begin{abstract}
Photometric reverberation mapping (RM) provides an efficient alternative to spectroscopic RM for probing the broad-line region (BLR) sizes in AGNs. In our previous work, we proposed the ICCF-Cut method, which extracts H$\alpha$ emission-line variability from broadband photometric light curves and measures the lags to estimate the BLR sizes. To further assess the reliability of this method, we conduct simultaneous photometric and spectroscopic monitoring of two nearby Seyfert galaxies, NGC 4151 and UGC 3374, over multiple observing seasons using the Lijiang 2.4 m telescope. By directly comparing the photometric and spectroscopic RM results in each season, we find that the extracted H$\alpha$ light curves using the ICCF-Cut method closely resemble those derived from spectroscopy. The photometric H$\alpha$ lags are also generally consistent with the spectroscopic H$\alpha$ lags within the uncertainties. For several seasons where the photometric H$\alpha$ lags are slightly underestimated, we find that the discrepancy may be caused by residual He\,\textsc{i} contamination in the extracted H$\alpha$ light curves. After correcting for this contamination, the photometric lag measurements become more consistent with the spectroscopic results. We further explore the performance of the ICCF-Cut method using the simulated light curves with different filter bandwidths. For our observational and simulated cases, the method can successfully recover the emission-line variability and lag measurements broadly consistent with the spectroscopic results. This provides further support for the applicability of the ICCF-Cut method in photometric RM.
\end{abstract}


\section{Introduction}\label{sec:intro}
Active galactic nuclei (AGNs) are among the most luminous sources of radiation in the Universe, powered by accretion onto supermassive black holes (SMBHs; \citealt{1969Natur.223..690L}). Although AGNs are known to play an important role in galaxy evolution \citep{2012ARA&A..50..455F,2013ARA&A..51..511K}, the key physical processes occur on scales that are too small to be spatially resolved with current facilities. Reverberation mapping (RM) provides an effective alternative by exploiting light-travel time delays between continuum variations and the responses of surrounding gas or dust \citep{1982ApJ...255..419B,1993PASP..105..247P}. This technique has become a powerful tool for probing black hole masses and the structure of accretion disks, broad-line regions (BLRs), and dusty tori in AGNs \citep{2021iSci...24j2557C}.

The most widely used method for black hole mass measurements in AGNs relies on the geometry and kinematics of the BLR. Therefore, spectroscopic reverberation mapping (SRM) is essential, as it enables the isolation of broad-line flux variations necessary for measuring the BLR response to continuum changes. The current SRM programs can be broadly classified into two categories. The first category focuses on specific types of AGNs or targeted scientific goals, such as LAMP \citep{2009ApJ...705..199B,2015ApJS..217...26B,2022ApJ...925...52U}, SEAMBH \citep{2014ApJ...782...45D,2015ApJ...806...22D,2016ApJ...825..126D,2018ApJ...856....6D}, and MAHA \citep{2018ApJ...869..142D,2020ApJ...905...77B,2022ApJS..262...14B}. These programs rely on dedicated monitoring of individual sources using slit spectrographs on 2 m-class or larger telescopes, making them observationally expensive and thus impractical to apply to large AGN samples. The second category is based on large spectroscopic surveys, such as SDSS-RM \citep{2015ApJS..216....4S,2019ApJS..241...34S,2024ApJS..272...26S,2017ApJ...851...21G,2019ApJ...887...38G,2019ApJ...880..126H,2020ApJ...901...55H} and OzDES-RM \citep{2021MNRAS.507.3771Y,2023MNRAS.522.4132Y,2022MNRAS.509.4008P,2025arXiv251201260P,2023MNRAS.520.2009M,2024MNRAS.531..163M,2025arXiv251201261M}. These programs use multi-object fiber spectroscopy to conduct RM for hundreds of AGNs. However, the limited cadence and signal-to-noise ratio generally result in relatively large uncertainties in lag measurements, and only a fraction of the monitored sources yield reliable lag detections.

Although previous SRM studies have established an empirical relation between the BLR size and the AGN continuum luminosity \citep{2013ApJ...767..149B}, enabling convenient black hole mass estimates from single-epoch spectra \citep{2020ApJ...903..112D}, direct BLR size measurements from SRM are still available for only about 200 AGNs, primarily at low redshift and moderate luminosity \citep{2024ApJS..275...13W,2024ApJS..272...26S}. Moreover, the large scatter and accretion-rate dependence of the BLR size–luminosity relation \citep{2019ApJ...886...42D,2026ApJ...998..311L} raise concerns about its applicability to low-luminosity and high-redshift AGNs. These limitations motivate the development of more efficient RM techniques that can be applied to larger and more complete AGN samples.

Photometric reverberation mapping (PRM) provides an alternative approach to tracing broad emission-line variability through photometric monitoring. Narrow-band and intermediate-band PRM techniques typically use dedicated filters to isolate the broad emission-line component and have been successfully applied in several RM campaigns \citep{2011A&A...535A..73H,2012A&A...545A..84P,2016ApJ...818..137J,2018A&A...620A.137R,2019ApJ...884..103K,2025ApJS..276...48S}. However, these approaches generally depend on specific filter configurations and redshift windows, limiting their application to large AGN samples. Broadband PRM enables RM studies based on large time-domain photometric surveys. Because broadband photometry is typically dominated by continuum emission, isolating the emission-line variability and obtaining reliable lag measurements remain challenging.

Several methods have been developed to recover emission-line lags from broadband photometry. \cite{2012ApJ...747...62C} introduced the cross-correlation function minus auto-correlation function (CCF-ACF) method to mitigate continuum contamination in broadband PRM. It assumes identical continuum variability across different bands or treats the emission-line flux ratio as an additional free parameter \citep{2012ApJ...756...73E,2013A&A...552A...1P,2013ApJ...769..124C}. \cite{2016ApJ...819..122Z} proposed the JAVELIN Pmap model for broadband PRM. This model characterizes AGN variability using a damped random walk process, describes the emission-line response with a top-hat transfer function, and recovers posterior lag distributions via Monte Carlo sampling. In our previous work \citep{2023ApJ...949...22M}, we proposed the ICCF-Cut method for broadband PRM, which incorporates single-epoch spectra to estimate the emission-line flux ratio and applies a “cut” procedure to extract the emission-line variability from broadband photometry.

Although the ICCF-Cut method has been applied to multi-epoch data from the Zwicky Transient Facility (ZTF; \citealt{2019PASP..131a8003M}) to perform broadband PRM for 23 AGNs \citep{2024ApJ...966....5M}, the reliability of the resulting photometric lags has not yet been directly validated. In previous studies, this reliability was primarily assessed by comparing the photometric H$\alpha$ lags with H$\beta$ lags reported from SRM in the literature. This comparison provided an initial validation of the ICCF-Cut method, but it suffers from two inherent limitations. First, the spectroscopic and photometric lags were generally measured at different epochs, whereas the emission-line lags are known to vary with the luminosity and accretion state of AGNs \citep{2020ApJ...903...51W,2022ApJS..263...10L,2023MNRAS.520.1807C,2024ApJ...976..176F}. Second, the comparison involved different emission lines, as the ICCF-Cut method extracts the H$\alpha$ response from broadband photometry but was validated against H$\beta$ lags, which are not expected to be identical \citep{2010ApJ...716..993B,2021ApJ...912...92F,2022ApJ...936...75L}.

\begin{figure*}[t]
    \centering
    \includegraphics[width=0.95\textwidth]{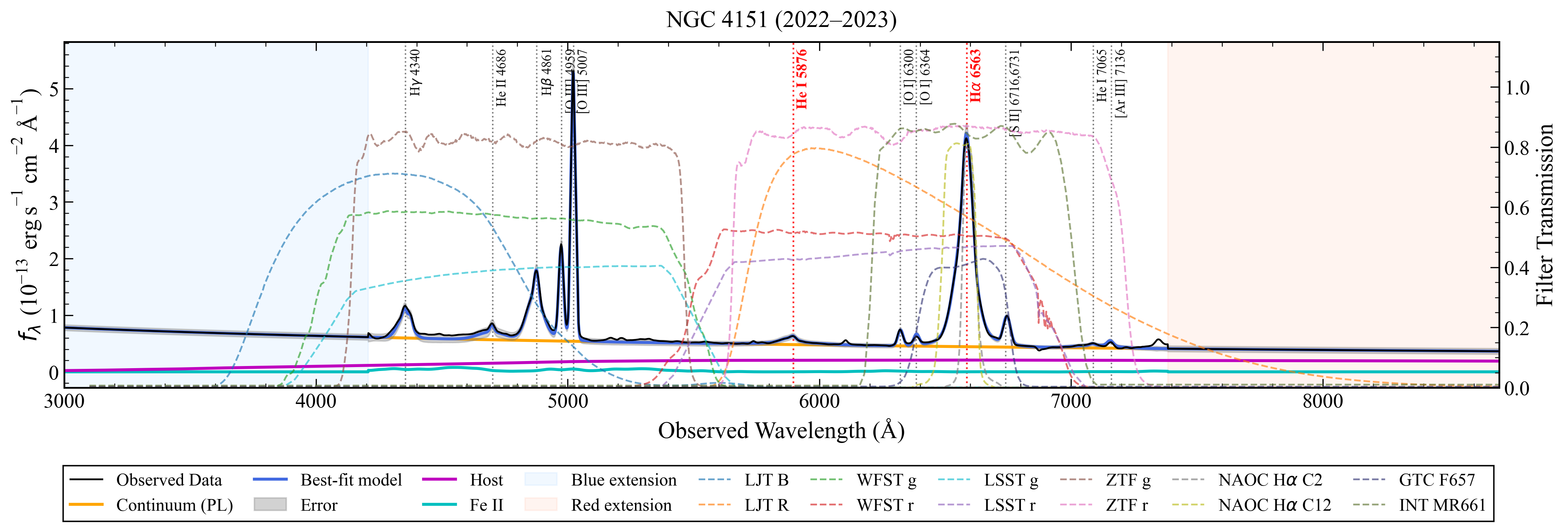} 
    \caption{Multicomponent fitting of the mean spectrum of NGC 4151 (2022–2023). The black solid line shows the observed spectrum. The blue solid line and gray shaded region depict the best-fit model and its uncertainties. The orange solid line denotes the AGN power-law continuum component, which is extrapolated toward the blue (blue shaded region) and red (red shaded region) ends. The cyan solid line represents the Fe\,\textsc{ii} template, and the purple solid line represents the host galaxy template. The vertical dotted lines mark the central wavelengths of various emission lines. The colored dashed lines indicate the transmission functions of different photometric filters.}
    \label{fig:spec}
\end{figure*}

Therefore, a definitive validation of the ICCF-Cut method requires a direct comparison between simultaneous PRM and SRM for the same emission line. In this work, we exploit multi-season observations of two Seyfert galaxies, NGC 4151 and UGC 3374, from the Lijiang 2.4 m telescope (LJT), which provides simultaneous multi-band photometry and spectroscopy. Previous SRM studies of these sources have revealed significant season-to-season variations in the H$\alpha$ lag \citep{2024ApJ...976..176F,2025ApJ...979..131F}. This provides an ideal opportunity to assess the reliability of the ICCF-Cut method. By performing a season-by-season comparison between photometric and spectroscopic H$\alpha$ lags and light curves, we can not only test the robustness of the method but also identify potential limitations and explore further refinements.

This paper is organized as follows. Section \ref{sec:obs} outlines the photometric and spectroscopic observations and data reduction. Section \ref{sec:method} introduces the methodologies adopted for PRM. Section \ref{sec:results} presents the lag measurements derived from PRM and evaluates their consistency with SRM. Section \ref{sec:discussion} discusses the reliability of the lag measurements and the potential systematic effects, as well as the validation of the ICCF-Cut method in PRM across different bandwidths. Finally, Section \ref{sec:summary} summarizes our main conclusions.

\begin{figure*}[t]
    \centering
    \includegraphics[width=0.90\textwidth]{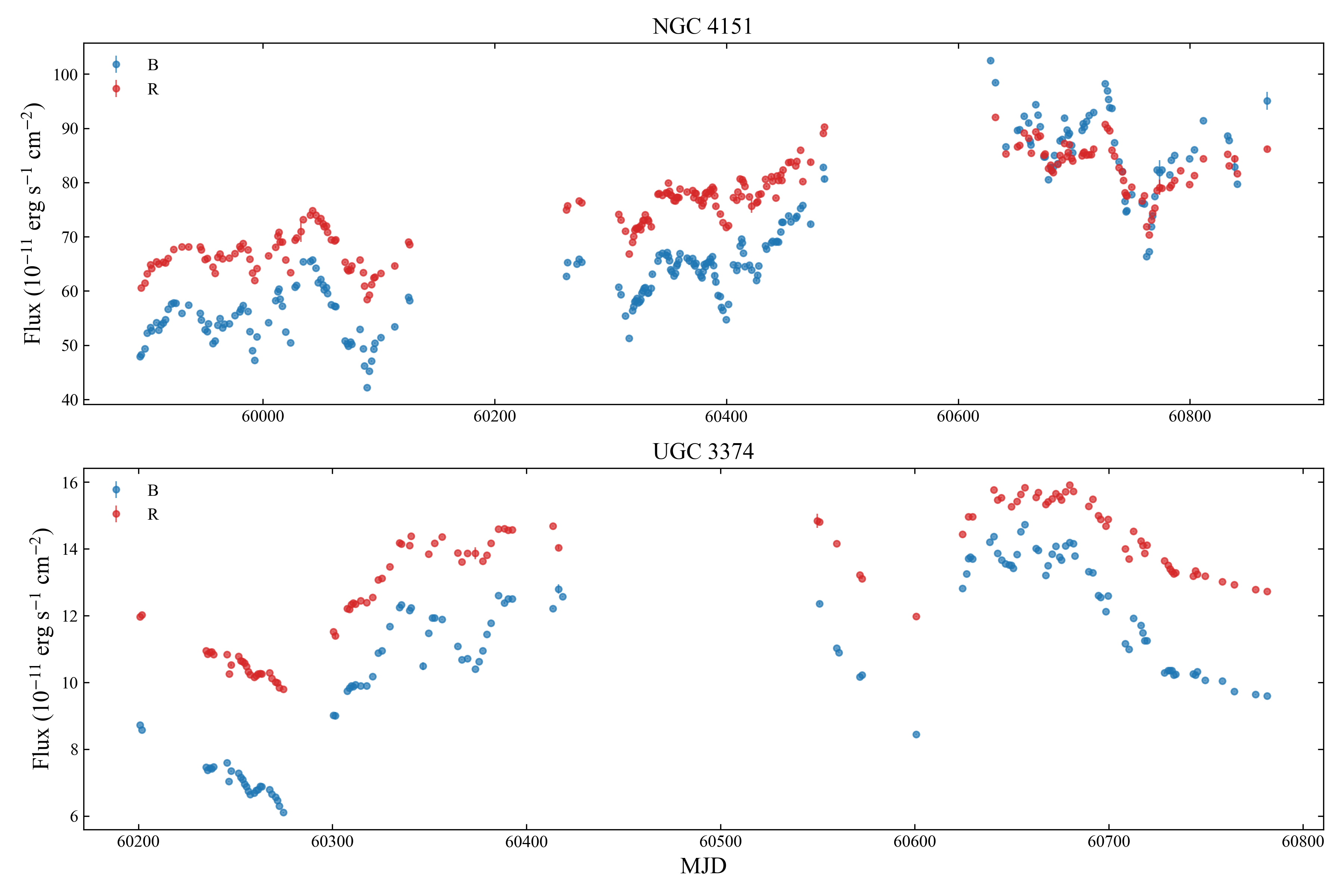} 
    \caption{Photometric light curves of NGC 4151 and UGC 3374 obtained from the LJT. Top panel: Photometric light curves of NGC 4151 during 2022–2025, where the red and blue points represent the $R$ and $B$ bands, respectively. Bottom panel: Photometric light curves of UGC 3374 during 2023–2025, with the same color scheme as the top panel.
    }
    \label{fig:lc_LJT}
\end{figure*}

\section{Observations and Data Reduction}\label{sec:obs}
We conducted simultaneous photometric and spectroscopic monitoring of two nearby Seyfert galaxies, NGC 4151 (z = 0.0033) and UGC 3374 (z = 0.0205), using the 2.4 m telescope at the Lijiang Observatory, Yunnan Observatories, Chinese Academy of Sciences. The telescope is equipped with the Yunnan Faint Object Spectrograph and Camera, which can rapidly switch between photometric and spectroscopic modes. The observations presented in this work span three observing seasons (2022–2025) for NGC 4151 and two seasons (2023–2025) for UGC 3374. Detailed descriptions of the observing strategy and data reduction procedures for the earlier seasons are provided in \citet{2022ApJ...936...75L,2024ApJ...976..176F,2025ApJ...979..131F,2026ApJ...997..326F}. For completeness, we briefly summarize the observational scheme and the data reduction steps below. 
\subsection{Photometric Monitoring}
For the photometric monitoring, Johnson $B$ and $R$ filters were used for both targets throughout all observing seasons. The raw images were processed using standard \texttt{IRAF} routines, including bias subtraction and flat-field correction. The instrument provides a field of view (FOV) of approximately $10^\prime \times 10^\prime$, allowing the selection of multiple comparison stars within the FOV for photometric calibration. Following \citet{2022ApJ...936...75L,2024ApJ...976..176F,2025ApJ...979..131F}, we conducted differential aperture photometry with an aperture radius of $4\farcs24$ (15 pixels). The background was estimated from an annulus with inner and outer radii of $14\farcs15$ and $19\farcs81$ (50 and 70 pixels). The size of the aperture was determined based on the FWHM of stellar profiles. The photometric uncertainties were estimated following the procedure described in \cite{2020ApJ...888...30F}, which has also been adopted in subsequent monitoring campaigns \citep{2022ApJ...936...75L,2024ApJ...976..176F,2025ApJ...979..131F}. Specifically, the uncertainty includes two components. The first component is the Poisson errors of the target and comparison stars. The second component is from the systematic uncertainties, which may be caused by the phase of the moon, weather conditions, etc. Detailed descriptions of the uncertainty estimation procedure can be found in \cite{2020ApJ...888...30F}. Finally, we obtained 243 $B$-band and 230 $R$-band epochs for NGC 4151, and 126 $B$-band and 112 $R$-band epochs for UGC 3374.
\subsection{Spectroscopic Monitoring}
For spectroscopic observations, we employed a long-slit mode to simultaneously obtain spectra for both the target and comparison star. During the 2022–2024 seasons for NGC 4151 and the 2023–2025 seasons for UGC 3374, we obtained spectroscopic observations using Grism~14 with a $5\farcs05$ slit and a UV-blocking filter. In the 2024–2025 season, we adopted a revised strategy for NGC 4151 and obtained spectra with both Grism~14 (without the UV-blocking filter) and Grism~8 \citep{2026ApJ...997..326F}. The flux calibration of the extracted spectra followed the standard procedure used in \citet{2024ApJ...976..176F,2025ApJ...979..131F}. We first selected spectra of a standard star obtained under clear weather conditions to determine the absolute flux of the comparison star. The calibrated spectra of the comparison star were then combined to construct a template spectrum. Finally, each nightly spectrum of the comparison star was compared to this template to derive a response function, which was subsequently applied to calibrate the target spectra. 

After flux calibration, the individual spectra obtained in each observing season were combined to generate a mean spectrum for both NGC 4151 and UGC 3374. These seasonal mean spectra were then subjected to detailed spectral decomposition using \texttt{PyQSOFit} \citep{2018ascl.soft09008G,2024ApJ...974..153R}. Prior to the fitting, each spectrum was corrected for Galactic extinction using the dust map of \citet{2011ApJ...737..103S}. The detailed fitting model includes several components: (1) a power-law function for AGN continuum; (2) a host-galaxy template from \citet{2018ApJ...866...92L}; (3) an Fe\,\textsc{ii} template from \citet{1992ApJS...80..109B}; (4) double-Gaussian profiles for the broad components of Balmer and Helium emission lines; and (5) single-Gaussian profiles for their narrow components as well as other common narrow forbidden lines. Figure \ref{fig:spec} illustrates an example of the resulting spectral decomposition for the seasonal mean spectrum of NGC 4151 during 2022–2023. For clarity, individual subcomponents of the emission lines are not explicitly labeled in the figure.

\subsection{Light Curves}
Consistent with our earlier work \citep{2024ApJ...976..176F,2025ApJ...979..131F}, we applied absolute flux calibration to the spectroscopic data, while only relative flux calibration was performed for the photometric data. To place the photometric light curves on a consistent flux scale, we convolved the seasonal mean spectrum with the transmission curves of the Johnson $B$ and $R$ filters to derive the mean flux levels for each observing season. The photometric light curves were then shifted to match the corresponding seasonal flux scales. To cover the full wavelength range of the photometric filters, each mean spectrum was extrapolated to 3000--8800\,\AA\ using the best-fit AGN power-law continuum. Although this extrapolation introduces some uncertainty into the derived mean flux levels, it affects only the absolute flux scale of the light curves, while the impact on their variability patterns is expected to be negligible. Since the subsequent lag analysis is based on the relative variability of the light curves rather than their absolute flux levels, the extrapolation has a negligible impact on the lag measurements. The final light curves of NGC 4151 and UGC 3374 are presented in Figure \ref{fig:lc_LJT}. NGC 4151 was monitored over three seasons (2022.11.10--2023.07.02; 2023.11.13--2024.06.24; 2024.11.13--2025.07.11), while UGC 3374 was monitored over two seasons (2023.09.13--2024.04.19; 2024.08.27--2025.04.17).

\section{Methodology}\label{sec:method}
\subsection{The ICCF-Cut Method}\label{subsec:ICCF-Cut}
The interpolated cross-correlation function (ICCF) is a widely used technique for measuring time lags between two light curves \citep{1987ApJS...65....1G,1994PASP..106..879W,1998PASP..110..660P}. However, it is often insufficient for PRM when the photometric band covering the emission line contains significant continuum contamination. To address this issue, we previously proposed the ICCF-Cut method \citep{2023ApJ...949...22M}, which combines the standard ICCF lag-measurement framework with a continuum-subtraction (“cut”) procedure. The key component of the method is the cut procedure, which incorporates single-epoch spectroscopic information to isolate the emission-line variability from broadband photometric light curves. In this framework, we define the photometric band with negligible line contamination as the continuum band, and the band containing both continuum and significant line emission as the line band. Using the single-epoch spectra, we estimate the continuum and emission-line contributions in the line band and subtract the continuum component. This process yields a “cut” light curve that predominantly traces the emission-line variability. The emission-line lag is then measured between the “cut” light curve and the continuum-band light curve using the standard ICCF approach.

In this work, we adopt the LJT $B$ band as the continuum band and the LJT $R$ band as the line band (including H$\alpha$ line). These two observational light curves can be written as: 
\begin{gather}
    F_B(t) \approx F_{B, \rm cont}(t), \label{eq:lcB} \\
    F_R(t) = F_{R, \rm line}(t) + F_{R, \rm cont}(t). \label{eq:lcR}
\end{gather}
To isolate the emission-line variability, we must accurately model and subtract $F_{R, \rm cont}(t)$. We assume that the continuum variability in the $R$ band follows the $B$-band variations, scaled by a flux ratio $\alpha$ and shifted by a continuum lag $\tau_{\rm cont}$. Then, the scaling factor $\alpha$ can be derived as:
\begin{equation}
\label{eq:alpha} 
\begin{aligned} 
\alpha &= \frac{F_{R, \rm cont}(t)}{F_{B, \rm cont}(t - \tau_{\rm cont})} \\
       &= \frac{F_{R}(t)-F_{R, \rm line}(t)}{F_{R}(t)} \times \frac{F_{R}(t)}{F_{B, \rm cont}(t - \tau_{\rm cont})} \\ 
       &\approx (1 - R_{\rm H\alpha}) \times \text{min}\left[ \frac{F_{R}(t)}{F_{B}(t)} \right]. 
\end{aligned} 
\end{equation}
For the first term, we estimate the H$\alpha$ ratio $R_{\rm H\alpha} \approx F_{R, \rm line}(t)/{F_{R}(t)}$ by convolving the seasonal mean spectrum with the LJT $R$ filter. In practice, the H$\alpha$ ratio varies slightly across different single-epoch spectra, but this approximation holds well over an observing season due to its minimal variation. The effect of the H$\alpha$ ratio on lag measurements is discussed in Section \ref{subsubsec:Sensitivity}. For the second term, we ignore the continuum lag since it is small relative to the emission-line lag and has a negligible effect on the lag measurements (see Section \ref{subsubsec:Sensitivity} for a detailed discussion). Following \cite{2023ApJ...949...22M}, we adopt the minimum value for the second term $F_{R}(t)/F_{B}(t)$. This conservative choice ensures that only the minimal plausible continuum contribution is removed, maximizing the retained H$\alpha$ signal in the resulting “cut” light curves. Finally, the extracted H$\alpha$ light curve in the $R$ band is expressed as:
\begin{equation}
\label{eq:lcline} 
F_{R, \rm line}(t) = F_{R}(t) - \alpha F_{B}(t).
\end{equation}
The uncertainty of the “cut” light curve is estimated as
$\Delta F_{\rm R,line}(t)=\Delta F_R(t)+\alpha \Delta F_B(t)$, 
where $\Delta F_R(t)$ and $\Delta F_B(t)$ are the photometric uncertainties of the $R$- and $B$-band light curves, respectively. This prescription provides a conservative estimate of the uncertainty of the “cut” light curve and accounts for potential correlated noise between the two photometric bands. It effectively considers an extreme case in the error propagation and thus helps avoid underestimating the uncertainties of the “cut” light curve.

While the “cut” light curve $F_{R, \rm line}(t)$ is dominated by H$\alpha$ emission, it may also include minor contributions from other emission lines such as He\,\textsc{i} (see Section \ref{subsubsec:HeI} for a detailed discussion). Consistent with our previous work \citep{2023ApJ...949...22M,2024ApJ...966....5M}, we also consider the effect of the host galaxy. Prior to the ICCF-Cut analysis, the host-galaxy component is estimated using the flux variation gradient method and subtracted from the observed light curves \citep{1981AcA....31..293C,1992MNRAS.257..659W,2012A&A...545A..84P}. After obtaining the “cut” light curve, we apply the ICCF to measure the photometric H$\alpha$ lag $\tau_{\rm cut}$ between the continuum-band light curve and the “cut” light curve. 

Based on the simultaneous SRM results, the expected H$\alpha$ lags for these five seasons are roughly around 0 to 30 days. We therefore uniformly set the lag search range from $-30$ to $60$ days to ensure that these expected lag scales are covered for all seasons. The interpolation step is set to 1 day. Following the standard procedures, the lag is defined as the centroid of the correlation coefficients above 80\% of the peak value. Uncertainties are estimated using the flux randomization and random subset sampling method with 1000 realizations \citep{2004ApJ...613..682P}. The resulting centroid lag distribution (CCCD) is constructed from these realizations. The final lag is taken as the median of the CCCD, with the 68\% credible interval adopted as the uncertainty. To ensure the robustness of lag measurements, we also performed sensitivity tests for each season using different search ranges and centroid thresholds, and found that the resulting lag measurements are not sensitive to these analysis settings.

\subsection{The JAVELIN Method}
We also employ a parallel technique, JAVELIN \citep{2011ApJ...735...80Z,2013ApJ...765..106Z,2016ApJ...819..122Z}, to estimate the lag for comparison. It models AGN variability as a damped random walk (DRW) process and relates different light curves through a transfer function, with model parameters inferred via Markov Chain Monte Carlo (MCMC) sampling. Since the photometric light curves contain both continuum and emission-line components, we adopt the Pmap model in JAVELIN. In this model, the continuum-band and line-band light curves are expressed as $f_c=c(t)+u_c$ and $f_l=\alpha\cdot c(t)+l(t)+u_l$, where $\alpha$ represents the scaling factor between the continuum variability in two bands. $u_c$ and $u_l$ denote constant flux contributions from the host galaxy. The continuum variability $c(t)$ and line variability $l(t)$ can be related by a top-hat transfer function centered on the time lag $\tau$ with width $\omega$ and amplitude $A$: 
\begin{gather}
l(t) = \int \Psi(t - t') c(t) \, dt', \\
\Psi(t) = \frac{A}{\omega} \quad \text{for} \quad \tau - \frac{\omega}{2} \leq t \leq \tau + \frac{\omega}{2}.
\end{gather}

Following the standard JAVELIN procedure, we first model the continuum light curve using a DRW model and obtain posterior constraints on the DRW parameters. These constraints are then used as priors in the subsequent lag analysis using the Pmap model.
For each JAVELIN run, we adopt the same lag search range as in the ICCF analysis, from $-30$ to $60$ days. We use the MCMC parameters $n_{chain}=n_{walkers}=n_{burn} = 100$ to ensure adequate sampling of the posterior distribution. The final lag $\tau_{\rm jav}$ is taken as the median of the posterior distribution, with the 1$\sigma$ uncertainty defined by the 16th and 84th percentiles.

\begin{figure*}[t]
    \centering
    \includegraphics[width=0.9\textwidth]{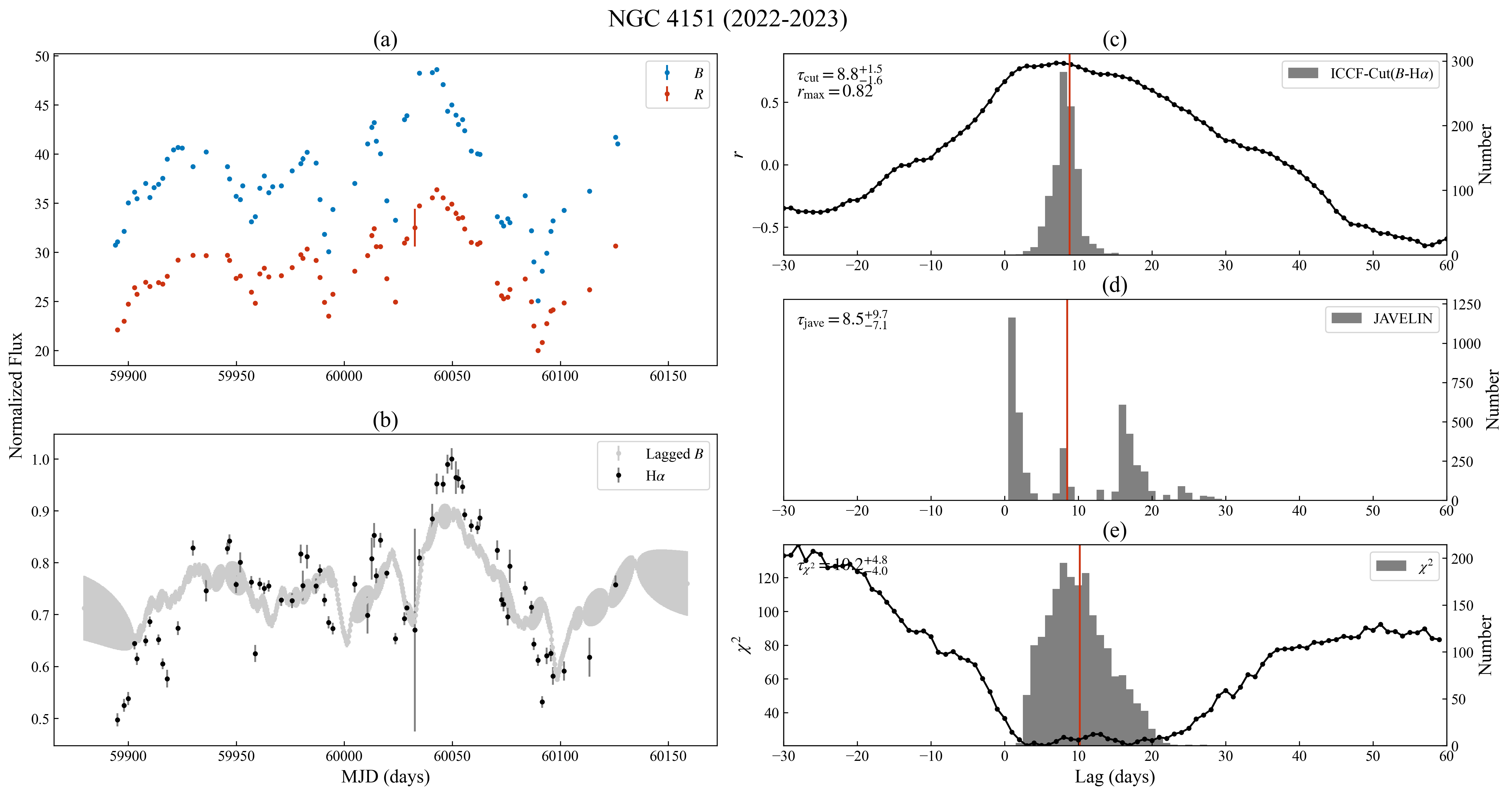} 
    \caption{Light curves and lag distributions for NGC 4151 (LJT 2022-2023). The top left panel shows the light curves of the continuum-band ($B$) and the line-band ($R$). The bottom left panel shows the extracted H$\alpha$ light curve compared with the lagged continuum-band light curve. The three right panels show the lag distributions between the continuum-band and extracted H$\alpha$ light curves with the ICCF-Cut, JAVELIN, and $\chi^{2}$ methods, respectively. The red line represents the median value of the lag distribution.}
    \label{fig:NGC4151_2223_result}
\end{figure*}

\subsection{The \texorpdfstring{$\chi^{2}$}{chi2} Method}
As noted in \citet{2023ApJ...949...22M}, the “cut” light curves typically exhibit larger flux uncertainties. To assess the impact of uncertainties and the feasibility of the PRM with large errors, we additionally employ the $\chi^2$ method \citep{2013A&A...556A..97C,2022ApJS..262...14B} as an alternative to the ICCF for measuring the photometric H$\alpha$ lag $\tau_{\chi^2}$ between the continuum-band light curve and the “cut” (H$\alpha$) light curve. The $\chi^2$ method employs the uncertainties to weight the data points in the light curves. For a given trial lag $\Delta t$, the $\chi^2$ statistic between the two light curves is calculated as
\begin{equation}
\chi^2(\Delta t) = \frac{1}{N}\sum_{i=1}^{N}
\frac{\left[x_i - A_{\chi^2} \, y_{i,\Delta t}\right]^2}
{\delta x_i^2 + A_{\chi^2}^2 \delta y_{i,\Delta t}^2},
\end{equation}
where $x_i$ and $y_{i,\Delta t}$ denote the continuum-band flux and the emission-line flux shifted by a lag $\Delta t$, respectively. $\delta x_i$ and $\delta y_{i,\Delta t}$ are their corresponding uncertainties. The normalization factor $A_{\chi^2}$ is formulated as 
\begin{equation}
A_{\chi^2} = \frac{S_{xy} + \left(S_{xy}^2 + 4S_{x3y}S_{xy3}\right)^{1/2}}{2S_{xy3}},
\end{equation}
where $S_{x3y} = \sum_{i = 1}^{N}x_i y_i\delta x_i^2$, $S_{xy3} = \sum_{i = 1}^{N}x_i y_i\delta y_i^2$, and $S_{xy} = \sum_{i = 1}^{N}\left(x_i^2\delta y_i^2 - y_i^2\delta x_i^2\right)$. By scanning over the same lag search range as in the ICCF and JAVELIN analyses, the time lag is identified as the value of $\Delta t$ that minimizes the $\chi^2$ function.

\section{Results}\label{sec:results}
\subsection{Lag measurement results}\label{subsec:Lag}
We perform photometric H$\alpha$ lag measurements using the ICCF-Cut, JAVELIN, and $\chi^2$ methods for NGC 4151 and UGC 3374 in each observing season. For convenience, the five observing seasons are hereafter denoted as NGC 4151 (2022–2023), NGC 4151 (2023–2024), NGC 4151 (2024–2025), UGC 3374 (2023–2024), and UGC 3374 (2024–2025). The lag measurement results for NGC 4151 (2022–2023) are displayed in Figure \ref{fig:NGC4151_2223_result} as an example. The remaining lag measurements are presented in Figure \ref{fig:NGC4151_UGC3374_LJT_result} in the Appendix. For each observing season, we measure the inter-band lag $\tau_{\rm inter}$ between the continuum-band (LJT $B$) and line-band (LJT $R$) light curves, as shown in Figure \ref{fig:iccf_interband}. In addition, we also collect simultaneous spectroscopic H$\alpha$ lags $\tau_{\rm ref}$ reported in the literature for comparison \citep{2024ApJ...976..176F,2025ApJ...979..131F}. All these lags are listed in Table \ref{tab:Lag}. In the following, we provide a detailed description of the lag measurement results for each observing season.

\begin{table*}[t]
\begin{center}
\caption{The lag measurements for the LJT light curves of NGC 4151 and UGC 3374 during different observing seasons.}
\label{tab:Lag}
\vspace{-10pt}
\setlength\tabcolsep{1.9mm}{
\begin{tabular}{cccccccccc}
\toprule
\toprule
Name&Season &$R_{\text{H}\alpha}$& $\tau_{\text{inter}}$/days  & $\tau_{\text{spec}}$/days & $\tau_{\text{ref}}$/days & $\tau_{\text{cut}}$/days  &$\tau_{\text{jav}}$/days & $\tau_{\chi^2}$/days & $\tau_{\text{cor}}$/days \\
(1) & (2)  &(3) &  (4) &(5)& (6) & (7) &(8)& (9)&(10)\\
\midrule
\multirow{3}{*}{NGC 4151} 
& 2022-2023  &0.2105& $1.71_{-0.73}^{+0.71}$ & $10.55_{-2.01}^{+1.93}$ & $10.65_{-2.15}^{+1.27}$ & $8.83_{-1.56}^{+1.51}$   &$8.51_{-7.07}^{+9.67}$ & $10.23_{-3.97}^{+4.82}$ & $9.34_{-1.88}^{+1.59}$ \\
& 2023-2024  &0.1925
& $1.27_{-0.68}^{+0.52}$ & $4.00_{-1.50}^{+1.53}$ & $3.77_{-1.27}^{+1.37}$ & $3.97_{-2.39}^{+2.50}$   &$3.44_{-1.99}^{+39.88}$ & $5.47_{-3.18}^{+5.53}$ & -- \\
& 2024-2025  &0.1899
& $1.76_{-0.67}^{+0.51}$ & -- & $21.06_{-9.90}^{+5.80}$ & $19.99_{-3.41}^{+3.95}$   &$28.10_{-25.02}^{+3.20}$ & $21.98_{-7.79}^{+9.00}$ & -- \\
\midrule
\multirow{2}{*}{UGC 3374} 
& 2023-2024  &0.1768
& $3.54_{-1.44}^{+1.01}$ & $25.50_{-2.01}^{+1.98}$ & $23.80_{-2.10}^{+2.05}$ & $20.07_{-2.49}^{+2.55}$   &$19.38_{-0.56}^{+0.13}$ & $20.59_{-6.02}^{+9.29}$ & $22.88_{-4.15}^{+2.57}$ \\
& 2024-2025  &0.1623& $4.21_{-5.22}^{+1.93}$ & -- & $38.63_{-3.50}^{+11.20}$ & $33.14_{-3.65}^{+3.17}$   &$21.61_{-14.07}^{+0.96}$ & $30.26_{-5.43}^{+4.88}$ & -- \\
\bottomrule
\end{tabular}}
\end{center}
\textbf{Note.} Column (1): Object name. Column (2): Observation season. Column (3): The H$\alpha$ ratio obtained by convolving the seasonal mean spectrum with the LJT $R$ filter. Column (4): The inter-band lag between the continuum-band (LJT $B$) and line-band (LJT $R$) light curves without ICCF-Cut processing. Column (5): The spectroscopic H$\alpha$ lag between the LJT $B$-band and spectroscopic H$\alpha$ light curves. Column (6): The reference spectroscopic H$\alpha$ lag reported in the literature. Columns (7)--(9): The photometric H$\alpha$ lag measured by the ICCF-Cut, JAVELIN, and $\chi^2$ methods, respectively. Column (10): The photometric H$\alpha$ lag measured by the ICCF-Cut method after correcting for the He\,\textsc{i} contamination. 

\end{table*}

\textbf{NGC 4151 (2022–2023)}: In Figure \ref{fig:NGC4151_2223_result}, the ICCF-Cut and $\chi^2$ methods yield consistent lag distributions that agree with the spectroscopic reference lag. The JAVELIN method also returns a similar lag measurement, but its lag distribution features two additional peaks around $\sim$1--2 days and $\sim$16--17 days. The smaller lag is likely caused by the continuum lag, as it closely matches our measured inter-band lag in this season. This value is also broadly consistent with previously reported continuum lags for NGC 4151 \citep{2025ApJ...986..137Z,2026ApJ...997..326F}. The larger lag may be related to the potential quasi-periodic variability in the light curves (see Section \ref{subsubsec:QPO} for further discussion).

\textbf{NGC 4151 (2023–2024)}: The result for this observing season is shown in the top-left panel of Figure \ref{fig:NGC4151_UGC3374_LJT_result}. The ICCF-Cut, JAVELIN, and $\chi^2$ methods provide consistent lag measurements. The lag distribution from the JAVELIN method shows several additional weaker peaks that lead to a large upper uncertainty. In addition, all three methods yield smaller H$\alpha$ lags than those from the previous observing season. Similarly, a smaller inter-band lag is also observed for this season. Importantly, these smaller photometric H$\alpha$ lag measurements are still in agreement with the simultaneous spectroscopic H$\alpha$ lag in this season.

\textbf{NGC 4151 (2024–2025)}: The result for this observing season is shown in the top-right panel of Figure \ref{fig:NGC4151_UGC3374_LJT_result}. The ICCF-Cut and $\chi^2$ methods yield consistent lag distributions, whereas the JAVELIN method delivers a larger, discrepant lag measurement. This discrepancy may be related to the potential quasi-periodic variability in the light curves (see Section \ref{subsubsec:QPO}). Despite this discrepancy, the reference spectroscopic H$\alpha$ lag remains broadly consistent with the photometric H$\alpha$ lags obtained from both the ICCF-Cut and $\chi^2$ methods.

\textbf{UGC 3374 (2023–2024)}: The result for this observing season is shown in the bottom-left panel of Figure \ref{fig:NGC4151_UGC3374_LJT_result}. The lag measurements from the ICCF-Cut, JAVELIN, and $\chi^2$ methods are highly consistent. However, they are slightly smaller than the simultaneous spectroscopic H$\alpha$ lag. This discrepancy is likely caused by additional contaminants in the extracted H$\alpha$ light curve, such as He\,\textsc{i} emission line or residual continuum component. We will discuss the impact of these contaminants in Section \ref{subsubsec:HeI} and further correct the lag measurements accordingly.

\textbf{UGC 3374 (2024–2025)}: The result for this observing season is shown in the bottom-right panel of Figure \ref{fig:NGC4151_UGC3374_LJT_result}. The ICCF-Cut and $\chi^2$ methods produce similar lag distributions. The lag distribution given by the JAVELIN method exhibits a primary peak accompanied by a weaker peak, potentially associated with the quasi-periodic features present in the light curves. Nevertheless, the photometric H$\alpha$ lags obtained from both the ICCF-Cut and $\chi^2$ methods remain smaller than the simultaneous spectroscopic H$\alpha$ lag in this season. A similar discrepancy has also been found for UGC 3374 (2023–2024), which is likely attributable to the residual contamination in the extracted H$\alpha$ light curve.

In summary, the photometric H$\alpha$ lags measured by the ICCF-Cut, JAVELIN, and $\chi^2$ methods are overall consistent with each other, as well as in broad agreement with the spectroscopic H$\alpha$ lags. The extracted H$\alpha$ light curves correlate well with the continuum-band light curves, with the maximum correlation coefficients exceeding 0.6 in all observing seasons. Compared to the JAVELIN method, the ICCF-Cut and $\chi^2$ methods yield more consistent results, as they share a similar methodology and use the same pair of light curves to measure the lags. However, the lags measured by the $\chi^2$ method typically have larger uncertainties due to its broader lag distribution. For the JAVELIN method, the lag distributions sometimes exhibit multiple peaks, leading to a skewed lag measurement with a large uncertainty. We also repeat the JAVELIN analysis multiple times for these observing seasons and find that the overall morphology of the posterior lag distributions remains largely unchanged between runs. In particular, sources exhibiting multi-peaked posterior distributions consistently retain similar multi-peaked structures, although the relative strengths of individual peaks may vary slightly between runs. Since the JAVELIN method assumes a specific AGN variability model and a simple transfer function, it may not fully capture the complexity of real light curves. Such multi-peak distributions often arise from sparse sampling, short monitoring duration, or quasi-periodic variability in the light curves \citep{2017ApJ...851...21G,2019ApJ...887...38G,2020ApJ...901...55H,2024ApJS..275...13W,2025ApJ...994..246W}. Finally, we adopt the ICCF-Cut measurements as the preferred photometric H$\alpha$ lags, while the results from the other two methods are used to assess the reliability of the ICCF-Cut measurements.

\begin{figure}[t]
    \centering
    \includegraphics[width=1\linewidth]{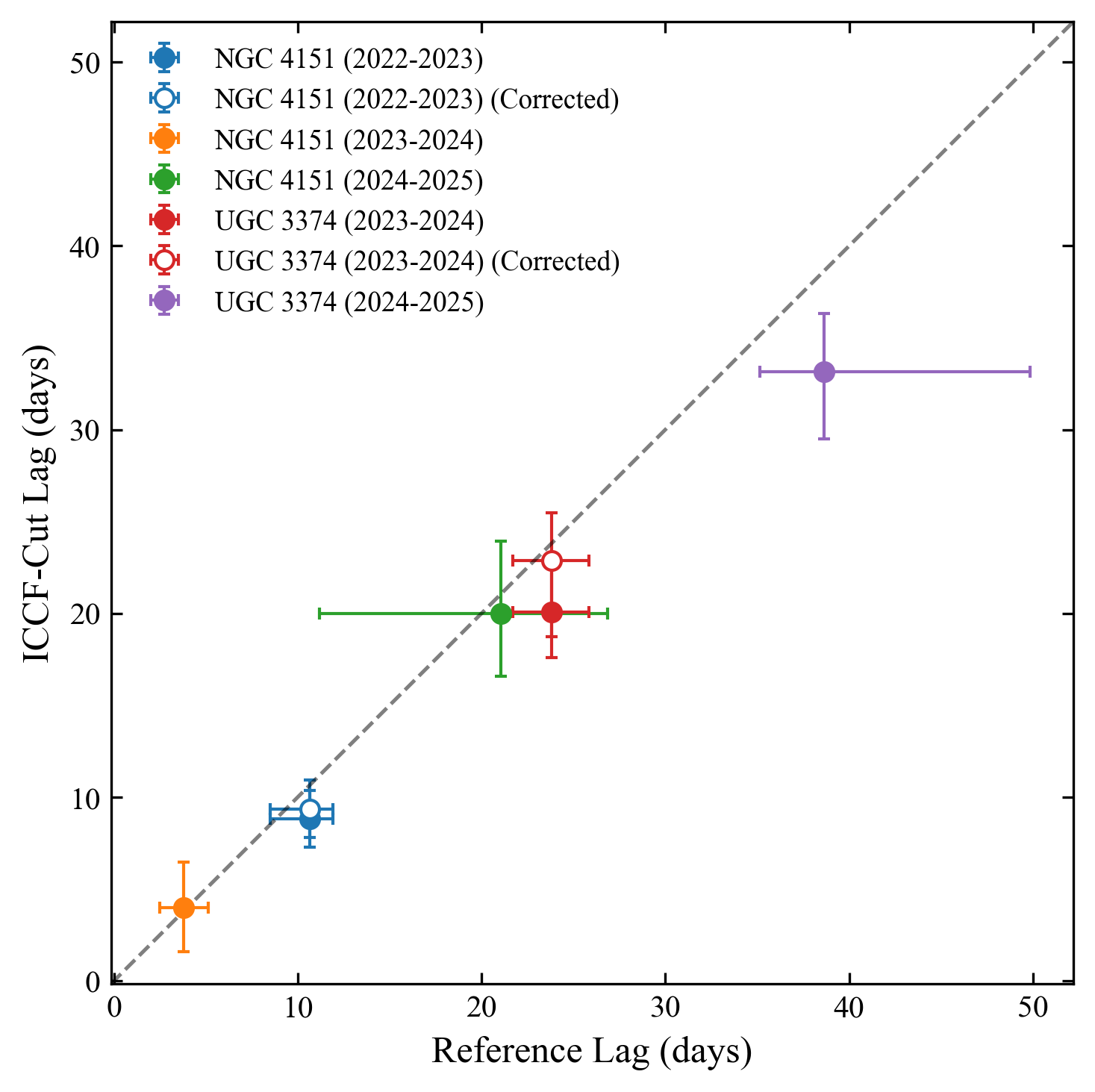} 
    \caption{Comparison between H$\alpha$ lags derived from the ICCF-Cut method and the reference lags reported in the literature. The solid points in different colors correspond to lags measured for NGC 4151 and UGC 3374 in different observing seasons, while the hollow points show the corrected ICCF-Cut lags accounting for He\,\textsc{i} contamination.}
    \label{fig:lag_compare}
\end{figure}

\subsection{Consistency Check with SRM}\label{subsec:check}

To further assess the reliability of the photometric H$\alpha$ lag measurements, we collect the simultaneous SRM results for consistency checks. For NGC 4151 (2022–2023), NGC 4151 (2023–2024), and UGC 3374 (2023–2024), \cite{2024ApJ...976..176F,2025ApJ...979..131F} performed detailed velocity-resolved RM analyses and provided the corresponding spectroscopic H$\alpha$ light curves and lag measurements. We also measure the lag ($\tau_{\rm spec}$) between the LJT $B$-band light curve and the spectroscopic H$\alpha$ light curve. As shown in the top three panels of Figure \ref{fig:iccf_spec_results}, these resulting lags ($\tau_{\rm spec}$) are in excellent agreement with the spectroscopic H$\alpha$ lags reported in the literature.
For NGC 4151 (2024–2025) and UGC 3374 (2024–2025), the spectroscopic H$\alpha$ light curves have not yet been published, as the related work is still in preparation. We therefore quote the corresponding spectroscopic H$\alpha$ lag measurements for comparison, as summarized in the sixth column of Table \ref{tab:Lag}. In Figure \ref{fig:lag_compare}, we compare the photometric H$\alpha$ lags $\tau_{\rm cut}$ with the spectroscopic H$\alpha$ lags $\tau_{\rm ref}$. Overall, the two sets of measurements are consistent within the uncertainties, although the photometric lags $\tau_{\rm cut}$ are generally slightly smaller than the corresponding spectroscopic lags $\tau_{\rm ref}$. This systematic offset may be caused by the residual continuum contamination arising from the conservative choice of adopting the minimum value of $F_{R}(t)/F_{B}(t)$ in Equation~\ref{eq:alpha}, which leaves a small residual continuum contribution in the extracted H$\alpha$ light curves. We also test adopting intermediate values of $F_{R}(t)/F_{B}(t)$ and find that the resulting lag measurements remain broadly consistent with the spectroscopic H$\alpha$ lags within the uncertainties. Additionally, the offset may also arise from residual He\,\textsc{i} contamination in the extracted H$\alpha$ light curves (see Section~\ref{subsubsec:HeI} for a detailed discussion).

For NGC 4151 (2022--2023), NGC 4151 (2023--2024), and UGC 3374 (2023--2024), where the published spectroscopic H$\alpha$ light curves are available, we perform the ICCF analysis to measure the compared lag $\tau_{\rm com}$ and the maximum correlation coefficient $r_{\rm max}$ between the extracted H$\alpha$ light curves and the spectroscopic H$\alpha$ light curves. The results for NGC 4151 (2022–2023), NGC 4151 (2023–2024), and UGC 3374 (2023–2024) are shown in Figure \ref{fig:Ha_Compare_All}. The compared lags for NGC 4151 (2022–2023) and NGC 4151 (2023–2024) are $0.4_{-1.9}^{+1.6}$ days and $1.4_{-2.3}^{+2.3}$ days, respectively, both of which are consistent with zero within the uncertainties. In contrast, the compared lag for UGC 3374 (2023–2024) is $6.1_{-2.3}^{+2.5}$ days. This positive lag suggests that the spectroscopic H$\alpha$ lag is systematically larger than the photometric H$\alpha$ lag, as also seen in Figure \ref{fig:lag_compare}. For all three observing seasons, the maximum correlation coefficients $r_{\rm max}$ exceed $0.7$, indicating a strong similarity between the extracted photometric H$\alpha$ light curves and the spectroscopic H$\alpha$ light curves. 

In summary, the agreement between the photometric and spectroscopic H$\alpha$ lag measurements and light curves for NGC 4151 (2022--2023), NGC 4151 (2023--2024), and UGC 3374 (2023--2024) supports the reliability of the ICCF-Cut method. For NGC 4151 (2024--2025) and UGC 3374 (2024--2025), the validation is currently limited to consistency with the spectroscopic lag measurements. Residual contamination in the extracted H$\alpha$ light curves may bias the lag measurements and should be carefully considered. 

\begin{figure*}[ht]
    \centering
    \includegraphics[width=0.9\textwidth]{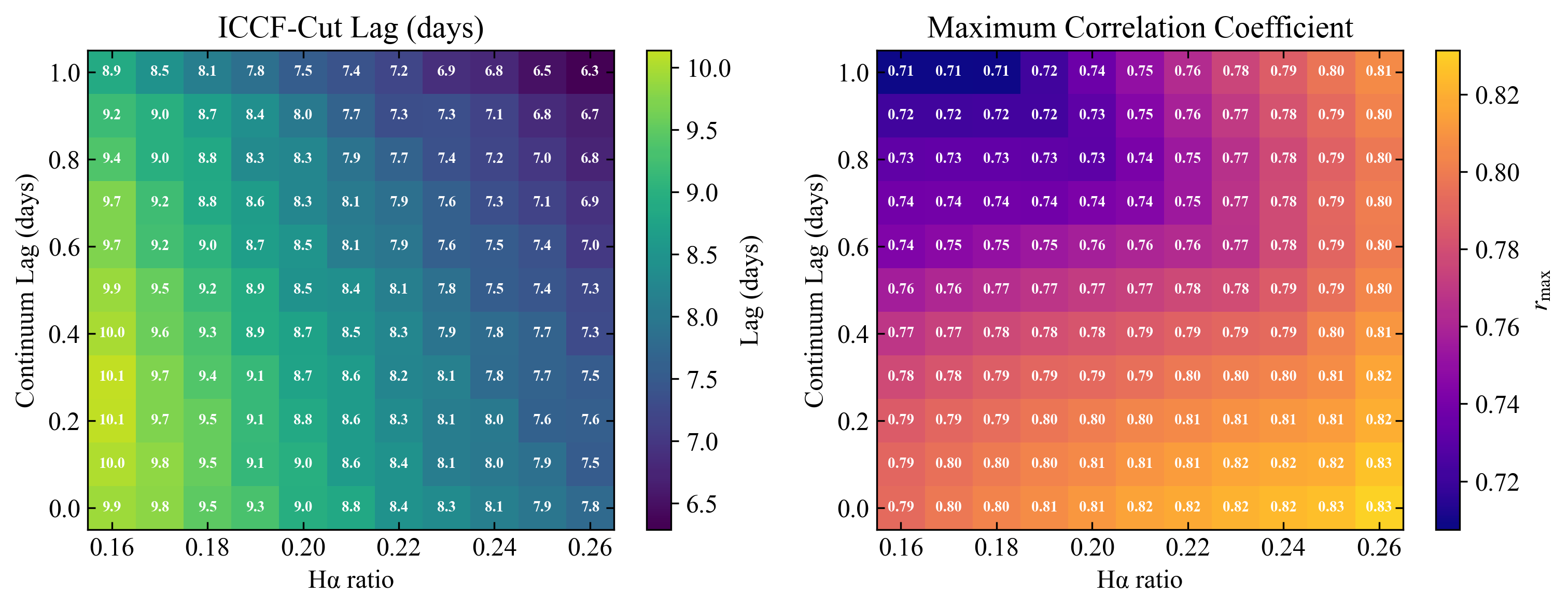} 
    \caption{Simulation results for NGC 4151 (LJT 2022–2023) illustrating the influence of the continuum lag and H$\alpha$ ratio on the ICCF-Cut measurements. Left panel: Heat map of the derived ICCF-Cut lag $\tau_{\rm cut}$, with colors denoting lag values and the corresponding numerical values labeled in white within each parameter cell. Right panel: Heat map of the maximum correlation coefficient $r_{\max}$, with colors indicating the correlation strength and white numbers showing the exact values.}
    \label{fig:iccf_heatmap}
\end{figure*}

\section{Discussion}\label{sec:discussion}
\subsection{Reliability and Systematic Effects in Lag Measurements}\label{subsec:Effect}
\subsubsection{Sensitivity to \texorpdfstring{H$\alpha$ }~Ratio and Continuum Lag}\label{subsubsec:Sensitivity}
The core of the ICCF-Cut method lies in extracting the H$\alpha$ emission component from the line-band light curves through a “cut” procedure. As shown in Equation \ref{eq:alpha}, this extraction process is influenced by the adopted H$\alpha$ ratio, $R_{\rm H\alpha}$, and the inter-band continuum lag, $\tau_{\rm cont}$. To evaluate the sensitivity of the method to these parameters, we take NGC~4151 (2022–2023) as a representative case and perform a series of simulations over different combinations of $R_{\rm H\alpha}$ and $\tau_{\rm cont}$. We first measure the H$\alpha$ ratio from each single-epoch spectrum of NGC 4151 (2022–2023) and find that the values range from 0.19 to 0.23. Compared to the value of $R_{\rm H\alpha} = 0.2105$ derived from the seasonal mean spectrum, no significant variation is observed over the observing season. To conservatively account for possible uncertainties, we therefore adopt a broader $R_{\rm H\alpha}$ range of 0.16–0.26 in the simulations. According to Equation~11 of \cite{2016ApJ...821...56F}, the standard thin-disk model predicts an inter-band lag of $0.21$ days between the LJT $B$ and $R$ bands. Observationally, continuum lags are typically larger than the predictions of the standard thin-disk model by a factor of a few \citep{2019ApJ...870..123E}. The physical origin of this discrepancy remains under debate, with proposed explanations including contributions from the BLR diffuse continuum \citep{2018ApJ...857...53C,2019MNRAS.489.5284K,2026ApJ...997..326F} or modifications to the accretion-disk model \citep{2018ApJ...854...93H,2019NatAs...3..251C,2020ApJ...891..178S,2021ApJ...907...20K}. A detailed discussion of this issue is beyond the scope of this work. Here we explore a conservative continuum-lag range of $0–1$ days, corresponding to approximately five times the thin-disk prediction.

\begin{figure*}[ht]
    \centering
    \begin{minipage}[b]{0.31\textwidth}
		\includegraphics[width=\linewidth]{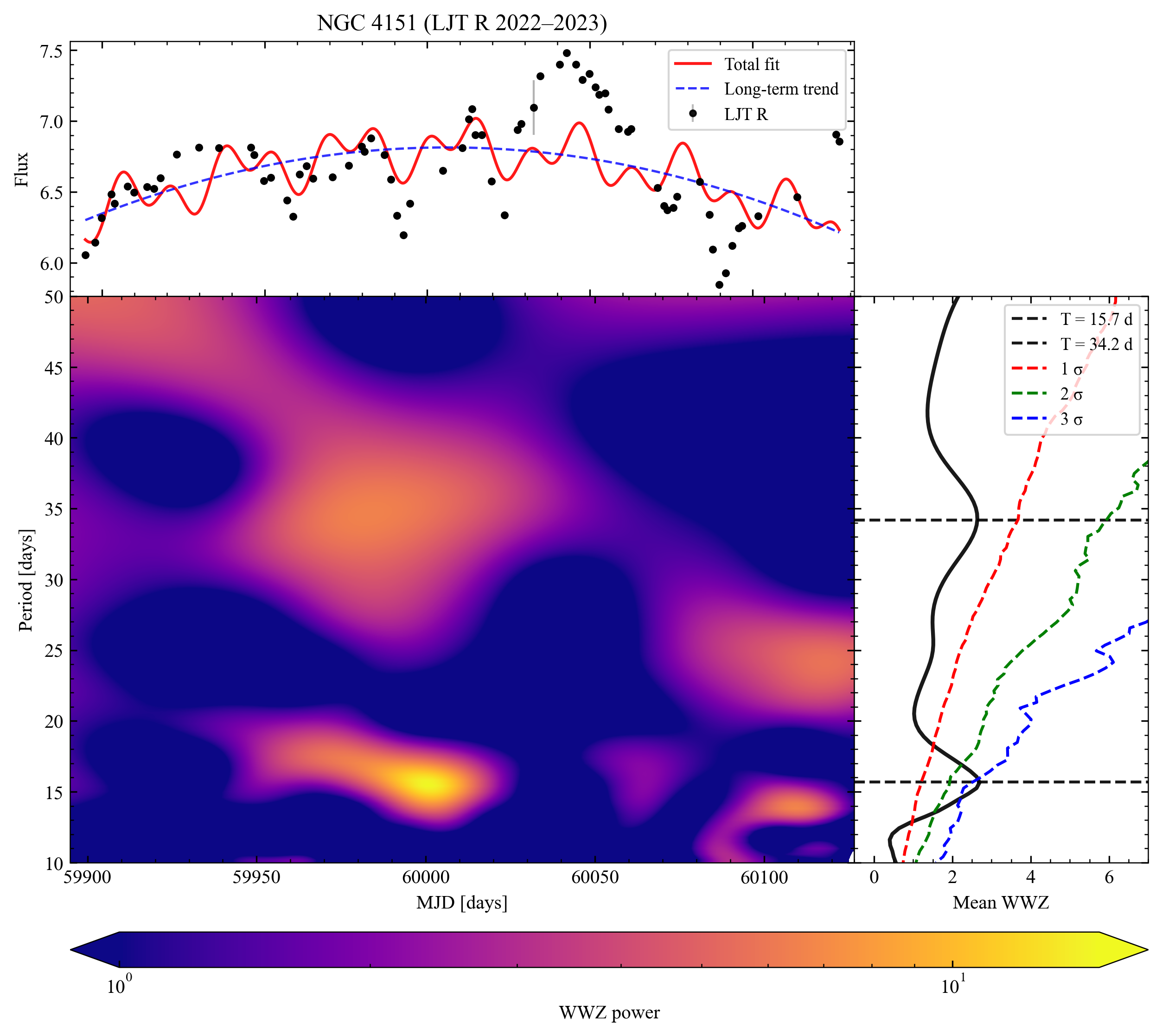}
	\end{minipage}%
	\begin{minipage}[b]{0.31\textwidth}
		\includegraphics[width=\linewidth]{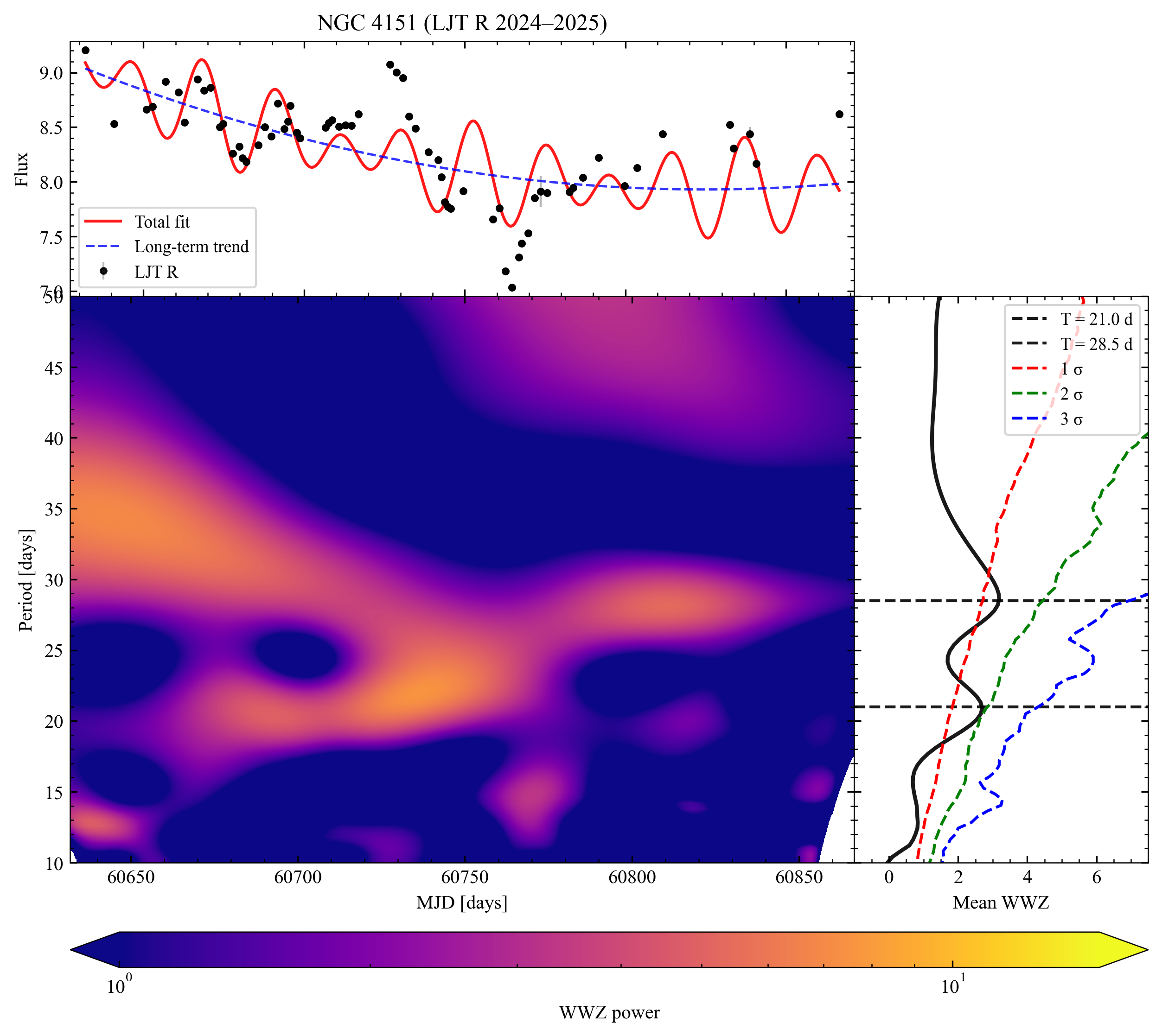}
	\end{minipage}%
    \begin{minipage}[b]{0.31\textwidth}
		\includegraphics[width=\linewidth]{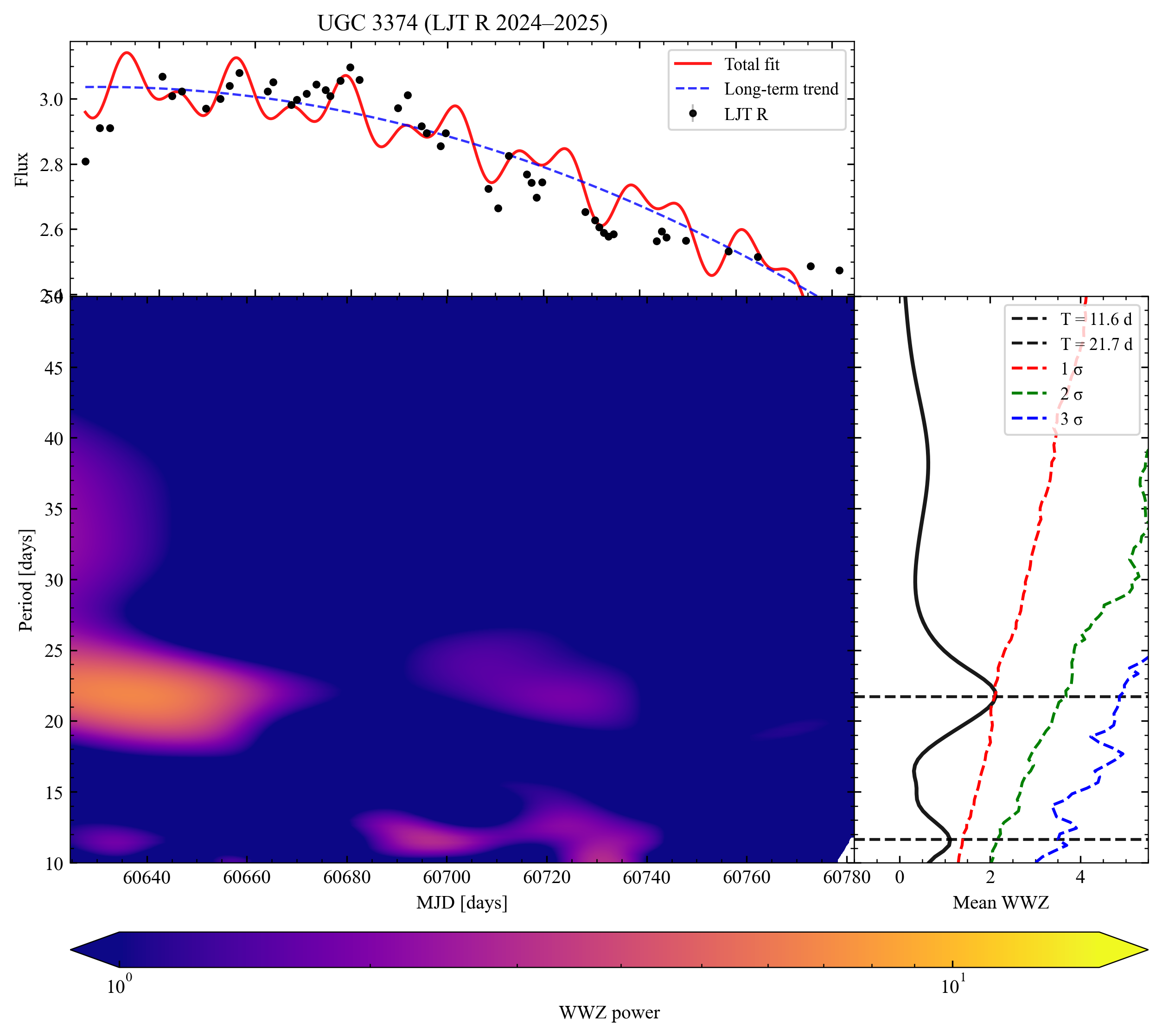}
	\end{minipage}%
    \caption{Left panel: Search for quasi-periodic variability in the LJT $R$-band light curve of NGC 4151 during 2022-2023. The top left subpanel shows the observed LJT $R$-band light curve (black points), the long-term trend fitted by a second-order polynomial (blue dashed line), and the fitted light curve considering the quasi-periodic component and the long-term trend (red solid line). The bottom left subpanel displays the 2D color contour of the WWZ power spectrum. The bottom right subpanel exhibits the time-averaged WWZ power (black solid line) and the 1$\sigma$, 2$\sigma$, and 3$\sigma$ confidence levels (orange, green, and red dashed lines, respectively). Middle panel: Same as the left panel, but for the LJT $R$-band light curve of NGC 4151 during 2024–2025. Right panel: Same as the left panel, but for the LJT $R$-band light curve of UGC 3374 during 2024–2025.}
    \label{fig:WWZ}
\end{figure*}

Based on the adopted parameter ranges, we construct a grid in which $R_{\rm H\alpha}$ varies from 0.16 to 0.26 with a step size of 0.01, and $\tau_{\rm cont}$ spans $0–1$ days with a step size of 0.1 days. This setup yields 100 simulated lag measurements, which are presented as heat maps in Figure \ref{fig:iccf_heatmap}. These simulations indicate that the measured photometric H$\alpha$ lag decreases as both the H$\alpha$ ratio and the continuum lag increase, while the corresponding maximum correlation coefficient increases. This trend can be qualitatively understood by considering Equations~\ref{eq:alpha} and~\ref{eq:lcline}. Adopting a larger H$\alpha$ ratio leads to an underestimation of the scaling factor $\alpha$ in Equation~\ref{eq:alpha}, causing the extracted H$\alpha$ light curve to retain a larger fraction of the continuum component. Since this residual continuum component exhibits a shorter lag and a stronger correlation with the LJT $B$-band light curve, it naturally biases the measured photometric H$\alpha$ lag toward smaller values and yields a higher maximum correlation coefficient. If an overestimated H$\alpha$ ratio is adopted in the analysis, it may partially account for the systematically smaller H$\alpha$ lags measured in some observing seasons. However, even when adopting the H$\alpha$ ratio derived from any single-epoch spectrum of NGC~4151 (2022–2023), where the values range from 0.19 to 0.23, the measured lag remains tightly constrained to 8.3–9.3 days. 

We perform the analogous tests for all observing seasons analyzed in this work and find similar behavior. The H$\alpha$ ratios measured from individual spectra within a given season typically vary by only about $\pm0.02$ around the value derived from the seasonal mean spectrum. Within these ranges, the resulting lag variations remain modest, and the recovered photometric H$\alpha$ lags remain consistent with the reference spectroscopic H$\alpha$ lags within the uncertainties. Based on these sensitivity tests, we estimate that the associated systematic uncertainty is typically of the order of $10\%$ in the measured lag for the datasets analyzed in this work. These results suggest that the limited variation of the H$\alpha$ ratio within a single observing season introduces only a modest bias in the measured photometric H$\alpha$ lag. Therefore, even when neglecting the limited seasonal variation of the H$\alpha$ ratio, we can still recover the lag measurements that are broadly consistent with the SRM results. The systematic effect resulting from omitting this time dependence is partially reflected in the final differences between the recovered photometric lags and the reference spectroscopic lags.

\subsubsection{Potential Influence of Quasi-periodic Variability on Lag Measurements}\label{subsubsec:QPO}
As noted in Section \ref{subsec:Lag}, the JAVELIN method sometimes produces lag posteriors with multiple peaks or yields lag measurements that differ from those obtained with other methods. Since different lag measurement techniques rely on different assumptions and strategies, they may respond differently to sampling quality, variability structure, and contamination in the light curves. The JAVELIN method assumes a damped random walk model together with a top-hat transfer function, which may make the inferred lag distributions sensitive to the validity of these assumptions and the quality of the light curves. In cases of a limited temporal baseline, local peaks or valleys in the light curves may align by chance. We further notice that the light curves associated with these discrepant lag measurements exhibit local repeating patterns on similar timescales. This suggests that the corresponding lag features may be related to the quasi-periodic variations in the light curves.

To further investigate this possibility, we perform a periodicity analysis on these light curves using the weighted wavelet Z-transform (WWZ) method \citep{1996AJ....112.1709F,2023ascl.soft10003K}. Compared with other methods, such as the autocorrelation function (ACF; \citealt{2013MNRAS.432.1203M}), the Lomb-Scargle Periodogram \citep{1976Ap&SS..39..447L,1982ApJ...263..835S} and the Jurkevich method \citep{1971Ap&SS..13..154J}, the WWZ method provides simultaneous localization in both time and frequency domains, enabling a more direct identification of localized periodic signals in the light curves. Following \cite{2024A&A...689A..35C}, we used the MCMC method to estimate the significance of the quasi-periodic signals in the WWZ map. We first characterize the observed light curves by calculating their power spectral density and probability distribution function. Based on these properties, we generate $10^4$ artificial light curves using the Emmanoulopoulos algorithm \citep{2013MNRAS.433..907E}. The WWZ analysis is then applied to each simulated realization, and the resulting distributions are used to derive the $1\sigma$, $2\sigma$, and $3\sigma$ significance level contours.

\begin{figure*}[ht]
    \centering
    \begin{minipage}[b]{0.9\textwidth}
        \includegraphics[width=\linewidth]{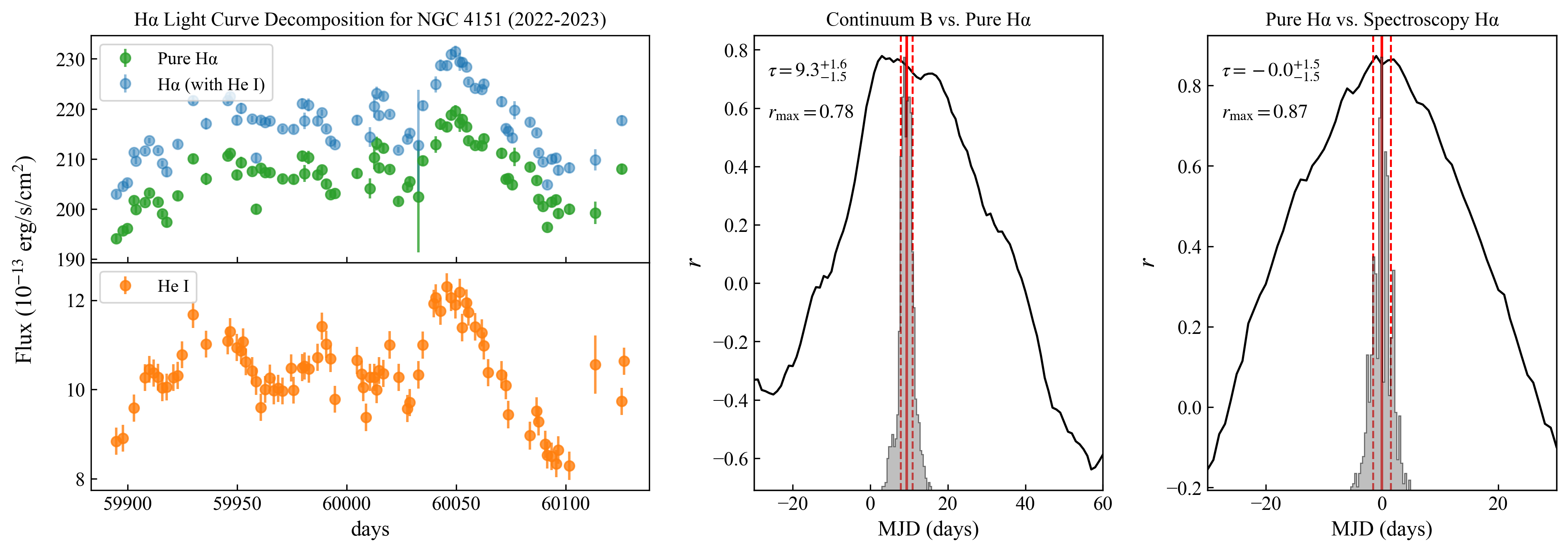}
        \vspace{-5mm} 
    \end{minipage}
    
    \begin{minipage}[b]{0.9\textwidth}
        \includegraphics[width=\linewidth]{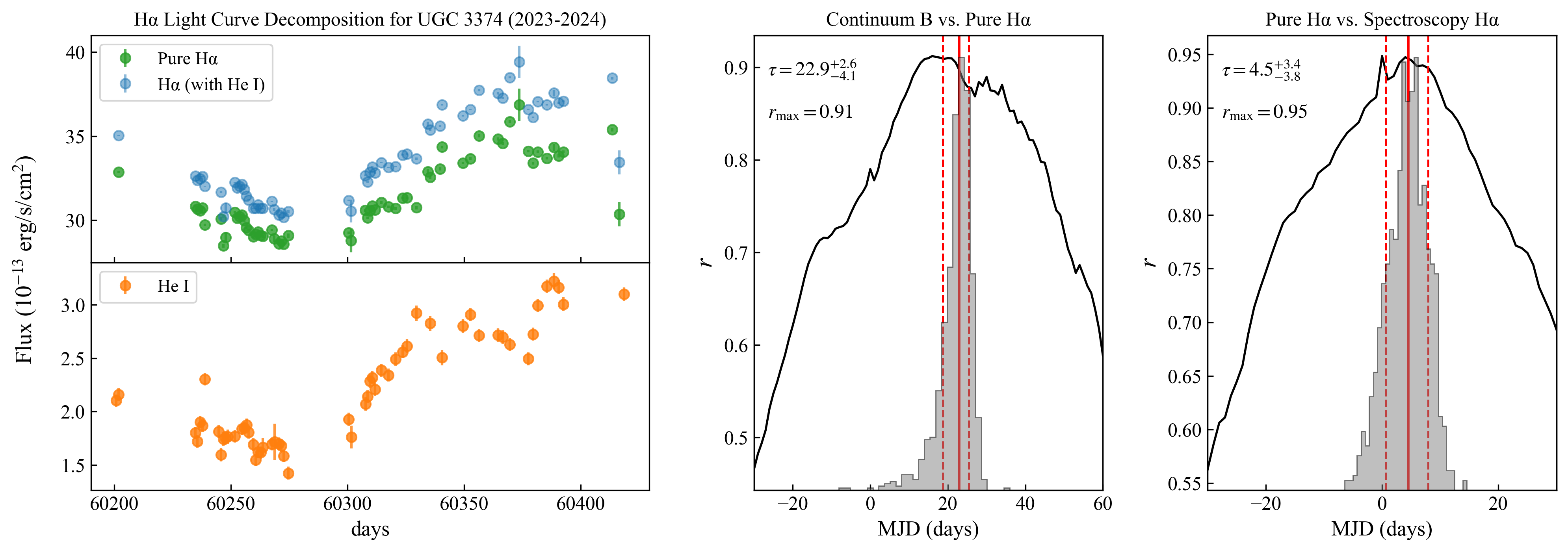}
    \end{minipage}
    
    \caption{Top panel: Decomposition of the H$\alpha$ light curves and corresponding lag measurements for NGC 4151 (2022–2023). The left subpanel shows the ICCF-Cut photometric H$\alpha$ light curve (green points), the spectroscopic He\,\textsc{i} light curve from \cite{2024ApJ...976..176F} (orange points), and the pure H$\alpha$ light curve after subtracting the He\,\textsc{i} contamination (green points). The middle subpanel presents the corrected ICCF-Cut lag between the LJT $B$-band light curve and the pure H$\alpha$ light curve. The right subpanel shows the corrected compared lag between the pure H$\alpha$ light curve and the spectroscopic H$\alpha$ light curve. For the middle and right subpanels, each panel exhibits the ICCF (black solid line) and the corresponding CCCD (gray shaded region). The red solid and dashed lines indicate the centroid lag and its $\pm1\sigma$ uncertainties, respectively. Bottom panel: Same as the top panel, but for UGC 3374 (2023–2024).}
    \label{fig:Ha_Decomp}
\end{figure*}

\begin{figure*}[ht]
    \centering
    \includegraphics[width=0.9\textwidth]{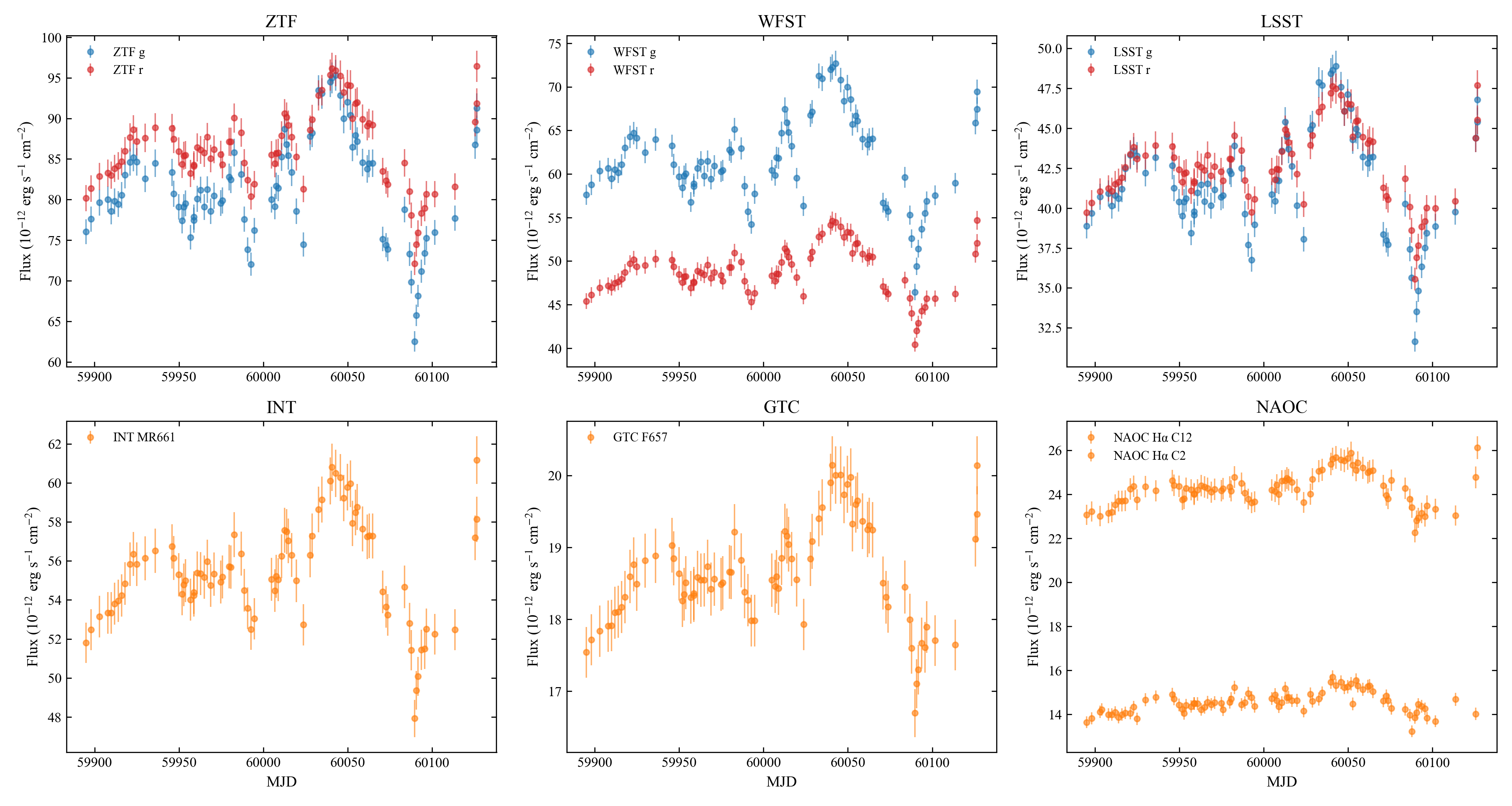} 
    \caption{Simulated light curves of NGC 4151 (2022–2023) across different photometric filter systems. Top panels: Broadband simulations for ZTF $g$/$r$ (left), WFST $g$/$r$ (middle), and LSST $g$/$r$ (right). Bottom panels: Narrow-band light curve simulations for INT MR661 (left), GTC F657 (middle), and NAOC H$\alpha$ C2/C12 filters (right).}
    \label{fig:lc_simulated}
\end{figure*}

For NGC 4151 (2022–2023), NGC 4151 (2024–2025), and UGC 3374 (2024–2025), we perform the WWZ analysis using the LJT $R$-band light curves as examples. The period search range is limited to 10–50 days, based on the sampling cadence and baseline of the light curves. The results are shown in Figure \ref{fig:WWZ}. For NGC 4151 (2022–2023), we identify a prominent periodic signal at 15.7 days with a significance exceeding the $3\sigma$ level. The characteristic timescale of this signal is broadly consistent with the additional peak around 16--17 days in the JAVELIN lag distribution shown in Figure~\ref{fig:NGC4151_2223_result}. This correspondence suggests that the quasi-periodic variability may contribute to the appearance of the additional lag peak. For NGC 4151 (2024–2025), the periodogram reveals two characteristic timescales at approximately 21 and 28.5 days, with the latter being similar to the JAVELIN lag measurement of $\tau_{\rm jav} = 28.10_{-25.02}^{+3.20}$ days. A similar case is found for UGC 3374 (2024–2025), where a periodic signal is detected at $\sim$21.7 days. This timescale is broadly consistent with the JAVELIN lag measurement of $\tau_{\rm jav} = 21.61_{-14.07}^{+0.96}$ days. However, the significance of the periodic signal only slightly exceeds the $1\sigma$ level and therefore provides only limited support for a possible connection between two timescales. To further assess the plausibility of the candidate periodicity, we model the observed light curves with sinusoidal functions corresponding to the detected periodic signals, combined with a second-order polynomial describing the long-term trend. As shown in Figure \ref{fig:WWZ}, the fitted light curves can reproduce some of the local quasi-periodic features present in the observations.

The above tests suggest that the quasi-periodic features identified in the light curves may contribute to the additional peaks observed in the JAVELIN lag distributions and thus serve as a possible source of the discrepancies among different lag measurement methods. On the other hand, these results also hint that, with appropriate modifications, lag measurement techniques may have the potential to serve as an alternative approach for periodicity searches, analogous to the transition from CCF to ACF. However, a more detailed discussion of this possibility is beyond the scope of this work.

\subsubsection{Impact of \texorpdfstring{He\,\scalebox{0.8}{I}}~ Contamination and Method Refinement}\label{subsubsec:HeI}
In Section \ref{subsec:ICCF-Cut}, we noted that the H$\alpha$ light curves extracted from the line band (LJT $R$) may still be contaminated by other emission. \citet{2026ApJ...997..326F} showed that the broad He\,\textsc{i} emission lines, although relatively weak in flux compared to H$\alpha$, still exhibit strong variability responses. Even a modest He\,\textsc{i} contribution within the LJT $R$ band may have a non-negligible impact on the measured photometric lags. As shown in the spectrum in Figure \ref{fig:spec}, we can clearly identify the broad He\,\textsc{i} emission line around 5876 \AA, which is within the range covered by the LJT $R$ filter. \cite{2024ApJ...976..176F,2025ApJ...979..131F} provided detailed He\,\textsc{i} light curves and lag measurements for NGC 4151 (2022-2023), NGC 4151 (2023-2024), and UGC 3374 (2023-2024). Below, we summarize the photometric H$\alpha$ lags in this work, along with the spectroscopic He\,\textsc{i} and H$\alpha$ lags reported in the literature. For NGC 4151 (2022-2023), the spectroscopic He\,\textsc{i} lag, spectroscopic H$\alpha$ lag, and photometric H$\alpha$ lag are $6.38_{-0.88}^{+0.91}$ days, $10.65_{-2.15}^{+1.27}$ days, and $8.83_{-1.56}^{+1.51}$ days, respectively. For NGC 4151 (2023-2024), these lags are $4.56_{-1.51}^{+0.98}$ days, $3.77_{-1.27}^{+1.37}$ days, and $3.97_{-2.39}^{+2.50}$ days, respectively. For UGC 3374 (2023-2024), they are $4.76_{-3.49}^{+1.20}$ days, $23.80_{-2.10}^{+2.05}$ days, and $20.07_{-2.49}^{+2.55}$ days, respectively. Although the photometric H$\alpha$ lags are consistent with the spectroscopic H$\alpha$ lags within the uncertainties, they systematically exhibit a small offset toward the spectroscopic He\,\textsc{i} lag. This systematic offset is likely a consequence of He\,\textsc{i} contamination within the LJT $R$ filter. We therefore introduce a refinement to the ICCF-Cut method to further correct for the contamination from He\,\textsc{i} emission.

We assume that the H$\alpha$ light curve contaminated by He\,\textsc{i} emission can be expressed as
\begin{equation}
L_{\rm mix}(t) = L_{\rm H\alpha}(t) + L_{\rm He\,\textsc{i}}(t),
\end{equation}
where $L_{\rm H\alpha}(t)$ and $L_{\rm He\,\textsc{i}}(t)$ denote the pure H$\alpha$ and He\,\textsc{i} light curves, respectively. Due to potential differences in flux calibration between spectroscopic and photometric observations, we introduce a flux scaling factor $S$ between the mixed light curve $L_{\rm mix}(t)$ and the ICCF-Cut H$\alpha$ light curve $L_{\rm cut}(t)$. This factor is defined as
\begin{equation}
S = \frac{F_{\rm H\alpha} + F_{\rm He\,\textsc{i}}}{\langle L_{\rm cut}(t) \rangle},
\end{equation}
where $F_{\rm H\alpha}$ and $F_{\rm He\,\textsc{i}}$ are the mean fluxes of the H$\alpha$ and He\,\textsc{i} emission lines obtained from \citet{2024ApJ...976..176F,2025ApJ...979..131F}, and $\langle L_{\rm cut}(t) \rangle$ is the mean flux of the ICCF-Cut H$\alpha$ light curve. Finally, by incorporating the He\,\textsc{i} light curves reported in \cite{2024ApJ...976..176F,2025ApJ...979..131F}, we can recover the pure H$\alpha$ light curve as
\begin{equation}
L_{\rm H\alpha}(t) = (F_{\rm H\alpha} + F_{\rm He\,\textsc{i}})\cdot
\frac{L_{\rm cut}(t)}{\langle L_{\rm cut}(t) \rangle} - L_{\rm He\,\textsc{i}}(t).
\end{equation}

\begin{table*}
\begin{center}
\caption{The lag measurements for the simulated light curves of NGC 4151 during 2022-2023.}
\label{tab:Lag_filters}
\vspace{-10pt}
\setlength\tabcolsep{1.6mm}{
\begin{tabular}{cccccccccc}
\toprule
\toprule
Name & Filter & Width/\AA & $R_{\text{H}\alpha}$ & $\tau_{\text{inter}}$/days & $\tau_{\text{spec}}$/days & $\tau_{\text{cut}}$/days  & $\tau_{\text{jav}}$/days & $\tau_{\chi^2}$/days & $\tau_{\text{cor}}$/days \\
(1) & (2) & (3) & (4) & (5) & (6) & (7)  & (8) & (9) & (10) \\
\midrule
\multirow{7}{*}{NGC 4151} 
 & ZTF r & 1417.53 & 0.2266 & $1.03_{-1.34}^{+0.55}$ & $10.47_{-2.09}^{+2.06}$ & $8.92_{-5.10}^{+5.43}$  & $17.35_{-12.68}^{+8.65}$ & $10.96_{-10.58}^{+9.11}$ & $9.68_{-6.67}^{+6.32}$ \\
 & WFST r & 1372.95 & 0.2323 & $0.99_{-0.96}^{+0.53}$ & $10.56_{-2.41}^{+2.05}$ & $8.22_{-4.83}^{+4.69}$  & $17.79_{-13.54}^{+9.28}$ & $9.49_{-8.96}^{+9.24}$ & $9.77_{-6.76}^{+6.06}$ \\
 & LSST r & 1206.92 & 0.2451 & $0.99_{-0.96}^{+0.55}$ & $10.50_{-2.29}^{+2.00}$ & $8.20_{-5.33}^{+4.32}$  & $16.80_{-12.54}^{+10.75}$ & $9.63_{-9.00}^{+9.24}$ & $9.42_{-6.88}^{+5.97}$ \\
 & INT MR661 & 767.79 & 0.3478 & $2.18_{-1.27}^{+0.96}$ & $10.55_{-2.01}^{+1.93}$ & $8.88_{-3.88}^{+3.19}$  & $15.53_{-10.09}^{+5.58}$ & $10.49_{-4.30}^{+5.05}$ & -- \\
 & GTC F657 & 351.04 & 0.4858 & $3.22_{-2.55}^{+1.30}$ & $10.55_{-2.01}^{+1.93}$ & $9.48_{-3.53}^{+3.52}$  & $15.48_{-6.56}^{+6.09}$ & $11.82_{-4.28}^{+4.96}$ & -- \\
 & NAOC H$\alpha$ C12 & 143.52 & 0.6199 & $4.93_{-3.41}^{+2.42}$ & $10.55_{-2.01}^{+1.93}$ & $10.53_{-3.98}^{+4.55}$  & $20.74_{-2.94}^{+2.60}$ & $14.37_{-5.52}^{+5.76}$ & -- \\
 & NAOC H$\alpha$ C2 & 64.13 & 0.4171 & $7.63_{-2.64}^{+2.57}$ & $10.55_{-2.01}^{+1.93}$ & $10.40_{-2.94}^{+3.14}$  & $10.60_{-1.52}^{+16.70}$ & $14.53_{-7.54}^{+8.90}$ & -- \\
\bottomrule
\end{tabular}}
\end{center}
\textbf{Note.} Column (1): Object name. Column (2): Filter used for simulating the line-band light curve. 
Column (3): Effective width of the filter (\AA). Column (4): The H$\alpha$ ratio obtained by convolving the seasonal mean spectrum with the line-band filter.
Column (5): The inter-band lag between the continuum-band and line-band light curves without ICCF-Cut processing. 
Column (6): The spectroscopic H$\alpha$ lag between the LJT $B$-band and spectroscopic H$\alpha$ light curves. 
Columns (7)--(9): The photometric H$\alpha$ lag measured by the ICCF-Cut, JAVELIN, and $\chi^2$ methods, respectively. 
Column (10): The photometric H$\alpha$ lag measured by the ICCF-Cut method after correcting the He\,\textsc{i} contamination.
\end{table*}

The close agreement among the spectroscopic He\,\textsc{i} lag, spectroscopic H$\alpha$ lag, and photometric H$\alpha$ lag for NGC 4151 (2023–2024) indicates that the He\,\textsc{i} contamination has an insignificant effect on the photometric H$\alpha$ lag measurement for this case. We therefore select NGC 4151 (2022–2023) and UGC 3374 (2023–2024) as representative cases to correct for He\,\textsc{i} contamination. For each case, we derive the pure H$\alpha$ light curves and measure the corrected lag, $\tau_{\rm cor}$, with respect to the continuum-band (LJT $B$) light curves, as well as the compared lag, $\tau_{\rm com}$, relative to the spectroscopic H$\alpha$ light curves. The results are shown in Figure \ref{fig:Ha_Decomp}. After correcting for He\,\textsc{i} contamination, the photometric H$\alpha$ lag for NGC 4151 (2022–2023) increases from 
$8.83_{-1.56}^{+1.51}$ days to $9.3_{-1.5}^{+1.6}$ days, making it more consistent with the spectroscopic H$\alpha$ lag of 
$10.65_{-2.15}^{+1.27}$ days. The compared lag is $0_{-1.5}^{+1.5}$ days with a high correlation coefficient of 0.87, 
indicating strong agreement between the spectroscopic and corrected photometric H$\alpha$ light curves. 
A similar improvement is found for UGC 3374 (2023–2024). The corrected photometric H$\alpha$ lag shifts from 
$20.07_{-2.49}^{+2.55}$ days to $22.9_{-4.1}^{+2.6}$ days, approaching the spectroscopic lag value of 
$23.80_{-2.10}^{+2.05}$ days. Meanwhile, the compared lag decreases from $6.1_{-2.5}^{+2.3}$ days to 
$4.5_{-3.5}^{+3.4}$ days, becoming closer to zero within the uncertainties. Overall, He\,\textsc{i} contamination introduces a modest bias in the photometric H$\alpha$ lag measurements. After correcting for this contamination using the spectroscopic He\,\textsc{i} light curves, the PRM results show better consistency with those from SRM. However, in the absence of spectroscopic constraints, directly correcting for He\,\textsc{i} contamination remains challenging. This result highlights that, although the ICCF-Cut method can provide reliable photometric lag measurements, achieving high-precision results may require source-specific contamination assessments and suitable spectroscopic priors.

\begin{figure*}[ht]
    \centering
    \includegraphics[width=0.8\textwidth]{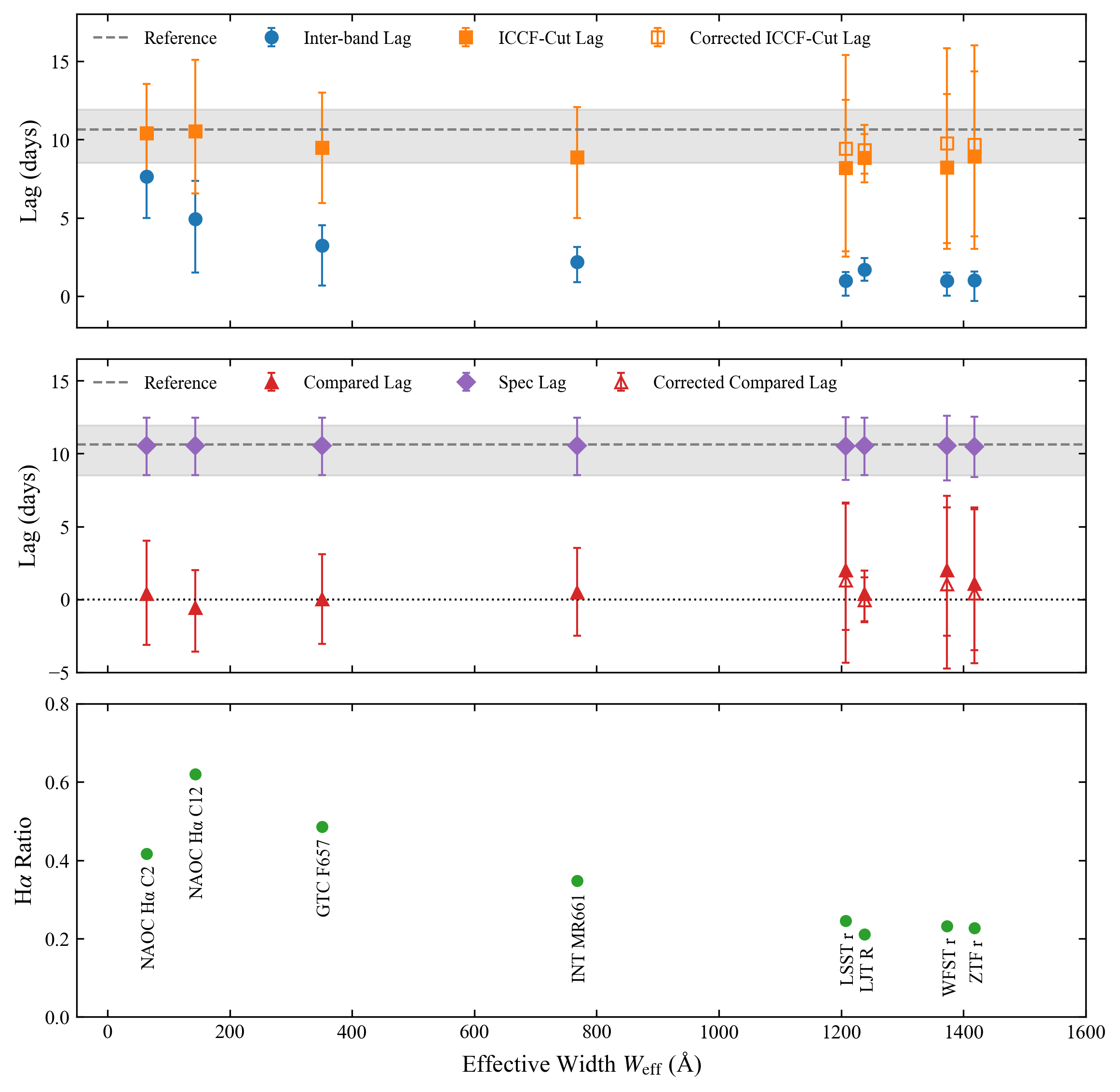} 
    \caption{PRM results for NGC 4151 (2022–2023) across different bandwidth filters. Top panel: Blue solid circles represent the inter-band lags measured directly from the continuum-band and line-band light curves without ICCF-Cut processing. Orange solid squares show H$\alpha$ lags measured from the ICCF-Cut method, while orange hollow squares denote ICCF-Cut lags corrected for potential He\,\textsc{i} contamination. Middle panel: Purple solid diamonds represent the H$\alpha$ lags between the spectroscopic H$\alpha$ light curves and the continuum-band light curves, including the observed LJT $B$-band and the simulated ZTF/WFST/LSST $g$-band light curves. Red solid triangles show the compared lags between the ICCF-Cut photometric H$\alpha$ light curves and the spectroscopic H$\alpha$ light curves, while red hollow triangles denote the corrected compared lags for He\,\textsc{i} contamination. Bottom panel: Green solid circles show the H$\alpha$ ratios for different bandwidth filters, with filter names labeled below each data point. For the top and middle panels, gray dashed lines and shaded regions indicate the spectroscopic H$\alpha$ lag and its uncertainties from \cite{2024ApJ...976..176F}; gray dotted lines denote zero lag.}
    \label{fig:lag_lambda}
\end{figure*}

\begin{figure*}[ht]
    \centering
    \includegraphics[width=0.9\textwidth]{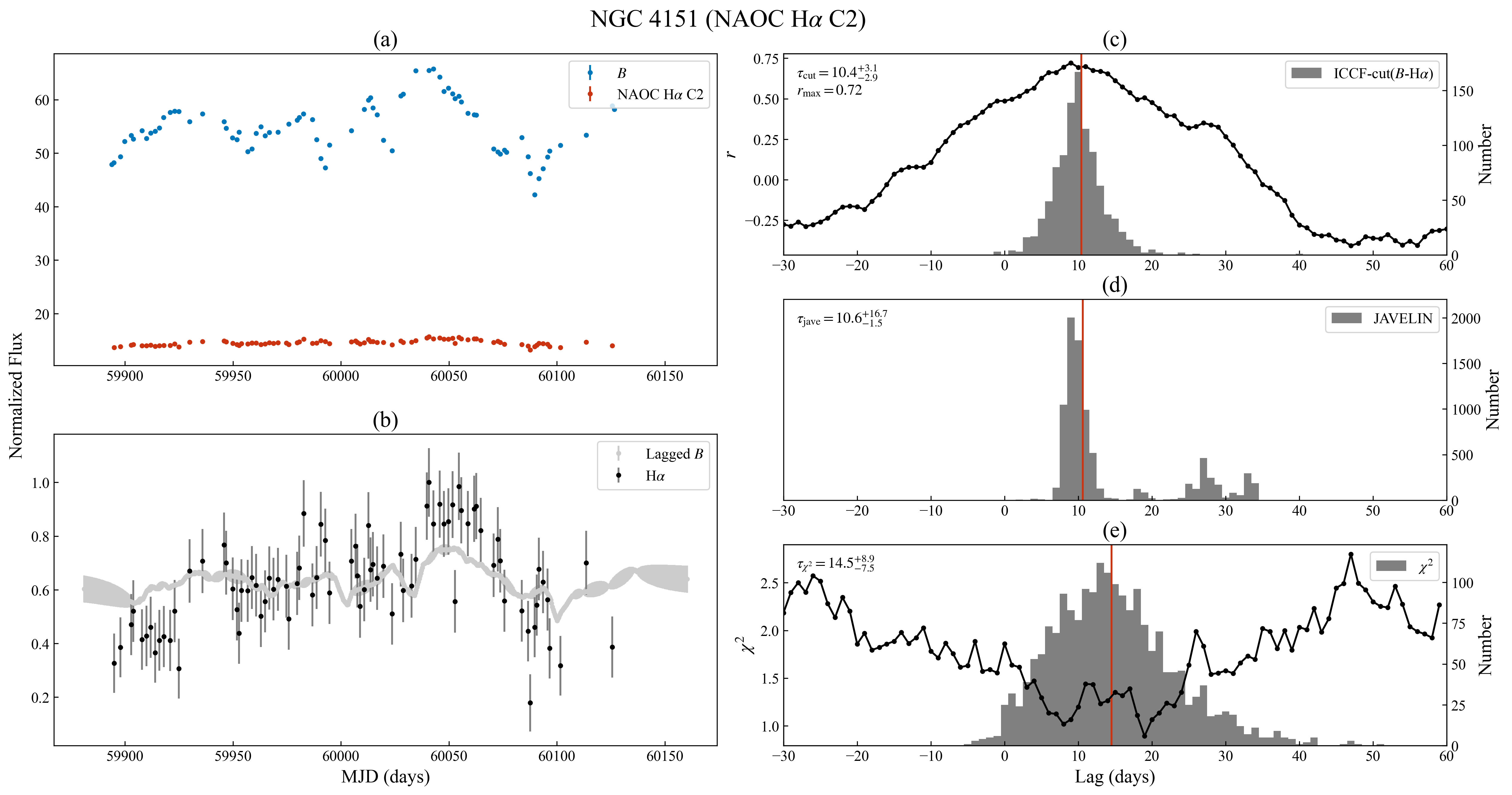} 
    \caption{Same as Figure \ref{fig:NGC4151_2223_result} but for NGC 4151 (NAOC H$\alpha$ C2 2022-2023).}
    \label{fig:NGC4151_NAOC_C2_result}
\end{figure*}

\subsection{Illustrative Application of the ICCF-Cut Method in PRM Across Different Bandwidths}\label{subsec:Further}
The previous results show that the ICCF-Cut method provides robust H$\alpha$ lag measurements in broadband PRM using the LJT $B$-band and $R$-band light curves. In this section, we present an illustrative single-case test to explore the performance of the method in PRM across different filter bandwidths, and therefore assess its potential application in future large multi-epoch photometric surveys. Taking NGC 4151 (2022–2023) as an example, we generate simulated light curves by convolving each single-epoch spectrum during this observing season with different photometric filters. To select filters with different bandwidths while ensuring coverage of the H$\alpha$ emission line, we adopted three broadband filters (ZTF $r$, WFST $r$, and LSST $r$), two intermediate-band filters (INT MR661 and GTC F657), and two narrow-band filters (NAOC H$\alpha$ C12 and NAOC H$\alpha$ C2) to construct the simulated line-band light curves. For broadband PRM, we used the corresponding ZTF $g$, WFST $g$, and LSST $g$ filters to construct the simulated continuum-band light curves. For the intermediate-band and narrow-band PRM cases, we retained the observed LJT $B$-band light curve as the continuum-band light curve. The transmission functions of these filters were obtained from the SVO Filter Profile Service\footnote{\url{https://svo2.cab.inta-csic.es/svo/theory/fps3/}} or provided by the respective instrument teams, and account for the effect of filter transmission efficiency, detector quantum efficiency, and optical system effects. These transmission functions are shown in Figure \ref{fig:spec}, and the corresponding simulated light curves are presented in Figure \ref{fig:lc_simulated}. 

Following the same procedure as described in Section \ref{subsec:Lag}, we first compute the H$\alpha$ ratios for different filters, which are listed in Table \ref{tab:Lag_filters}.
The bottom panel of Figure \ref{fig:lag_lambda} shows a clear decreasing trend of the H$\alpha$ ratios with increasing effective width of the filters.
The relatively low H$\alpha$ ratio for the filter with the smallest width (NAOC H$\alpha$ C2) is mainly caused by the broad H$\alpha$ component being slightly offset from the transmission window of this narrow-band filter. Before performing the PRM analysis, we measure the inter-band lags $\tau_{\rm inter}$ between the simulated continuum-band and line-band light curves (see Figure \ref{fig:iccf_interband}), as well as the spectroscopic lags $\tau_{\rm spec}$ between the simulated continuum-band light curves and the spectroscopic H$\alpha$ light curves (see Figure \ref{fig:iccf_spec_results}).
The results are listed in Table \ref{tab:Lag_filters}, and Figure \ref{fig:lag_lambda} further illustrates the relationship between these lags and the effective widths of filters. The measured $\tau_{\rm spec}$ values are consistent across different filters and agree with the spectroscopic H$\alpha$ lag from \cite{2024ApJ...976..176F}. The inter-band lags, $\tau_{\rm inter}$, decrease with increasing effective width due to stronger continuum contamination in broader filters. This anti-correlation explains why narrow-band or intermediate-band filters are commonly adopted in previous PRM studies to trace emission-line variability \citep{2025ApJS..276...48S,2019ApJ...884..103K,2017PASP..129i4101P,2012A&A...545A..84P}. 

However, we note that even with an appropriate narrow-band filter, the lag measured directly from photometric light curves remains significantly smaller than the reference spectroscopic H$\alpha$ lag when the continuum contamination is not corrected. We therefore apply the ICCF-Cut, JAVELIN, and $\chi^2$ methods to the simulated light curves to measure the photometric H$\alpha$ lags. As an example, Figure \ref{fig:NGC4151_NAOC_C2_result} presents the lag measurements derived from the light curves simulated using the narrowest filter (NAOC H$\alpha$ C2). All three methods yield consistent lag measurements within the uncertainties, with $\tau_{\rm cut} = 10.4_{-2.9}^{+3.1}$ days, $\tau_{\rm jav} = 10.6_{-1.5}^{+16.7}$ days, and $\tau_{\rm \chi^2} = 14.5_{-8.9}^{+7.5}$ days. After correcting for continuum contamination using the ICCF-Cut method, the resulting photometric H$\alpha$ lag is significantly larger than the inter-band lag, $\tau_{\rm inter} = 7.6_{-2.6}^{+2.6}$ days, and becomes highly consistent with the reference spectroscopic H$\alpha$ lag, $\tau_{\rm ref} = 10.65_{-2.15}^{+1.27}$ days. This result suggests that, for the NGC 4151 (2022-2023) dataset considered here, the ICCF-Cut method also performs well in narrow-band PRM.

For the remaining simulated light curves, we performed similar analyses. The corresponding lag measurements are listed in Table \ref{tab:Lag_filters} and displayed in Figure \ref{fig:NGC4151_simulated_result} in the Appendix. For all simulated cases, the ICCF-Cut method returns lag measurements consistent with the reference spectroscopic H$\alpha$ lag, with $\tau_{\rm cut}$ ranging from 8 to 11 days and maximum correlation coefficients above 0.7. The $\chi^2$ method also yields comparable lags around 10 days but with larger uncertainties. In contrast, the JAVELIN method produces systematically larger lag measurements around 15–20 days. As discussed in Section \ref{subsubsec:QPO}, these discrepant JAVELIN lags are likely caused by a highly significant 15.7-day periodic signal present in the light curves. In addition, we compare the photometric H$\alpha$ light curves extracted by the ICCF-Cut method with the spectroscopic H$\alpha$ light curves and measure the corresponding compared lags (see Figure \ref{fig:Ha_Compare_All}).

To more clearly illustrate the performance of the ICCF-Cut method in PRM with different filter bandwidths, Figure \ref{fig:lag_lambda} further presents the dependence of the ICCF-Cut lags and the compared lags on the filter bandwidths. For all simulated cases, the ICCF-Cut method significantly improves the lag measurements compared to the inter-band lags obtained directly without continuum contamination correction. The resulting H$\alpha$ lags are consistent with those from SRM, and the compared lags are near zero within the error bars. In comparison with broadband PRM, the ICCF-Cut lags in intermediate-band and narrow-band PRM are more consistent with the reference spectroscopic H$\alpha$ lag, and the corresponding compared lags are closer to zero. For broadband PRM, although the photometric H$\alpha$ lag has improved significantly, it remains slightly smaller than the reference spectroscopic H$\alpha$ lag. We attribute this difference to the residual continuum and He\,\textsc{i} contamination in the extracted H$\alpha$ light curves. This is also evident in the spectrum shown in Figure \ref{fig:spec}, where the transmission of broadband filters covers the He\,\textsc{i} emission line, whereas the intermediate-band and narrow-band filters avoid this contamination. We further correct the He\,\textsc{i} contamination in broadband PRM results and compute the corrected ICCF-Cut lags and compared lags, which are listed in Table \ref{tab:Lag_filters} and shown in Figure \ref{fig:lag_lambda}. After the correction, the ICCF-Cut lags are in better agreement with the reference value, with the compared lags approaching zero. 

Overall, this single-case experiment suggests that the ICCF-Cut method can mitigate continuum contamination and recover H$\alpha$ lag measurements in PRM across different bandwidths at least for the NGC 4151 (2022–2023) dataset. Notably, the method yields lag measurements more consistent with SRM in narrow-band PRM, while the improvement in continuum contamination correction is more significant in broadband PRM. Although these results are encouraging, this analysis serves primarily as an illustrative case study based on a single object and monitoring season. A more comprehensive assessment of the method across a wider range of AGN properties and observing conditions will be necessary in our future work.

\section{Summary}\label{sec:summary}
In this work, we assess the performance of the ICCF-Cut method through comparisons between simultaneous photometric and spectroscopic RM results obtained over multiple observing seasons for NGC 4151 and UGC 3374. The main results are summarized as follows:

\begin{enumerate}
    \item For each observing season of NGC 4151 and UGC 3374, the ICCF-Cut method successfully extracts the H$\alpha$ emission-line variability from the LJT $R$-band light curves, and yields robust lag measurements with high maximum correlation coefficients ($R_{\rm max} > 0.6$). These ICCF-Cut lags are generally consistent with those derived from the JAVELIN and $\chi^2$ methods. By comparison with simultaneous SRM results, most photometric H$\alpha$ lags are consistent with the spectroscopic H$\alpha$ lags within the uncertainties. For NGC 4151 (2022--2023), NGC 4151 (2023--2024), and UGC 3374 (2023--2024), where spectroscopic H$\alpha$ light curves are available, the cross-correlation analyses between the extracted photometric H$\alpha$ light curves and the spectroscopic H$\alpha$ light curves yield near-zero compared lags and high correlation coefficients ($R_{\rm max} > 0.7$). These results further support the applicability of the ICCF-Cut method for mitigating continuum contamination and recovering H$\alpha$ emission-line variability for the sources analyzed in this work.
    
    \item For NGC 4151 (2022--2023), we further investigate the reliability and potential systematic effects of the ICCF-Cut lag measurements through simulations covering a range of H$\alpha$ ratios and continuum lags. The results show that larger H$\alpha$ ratios and longer continuum lags tend to produce slightly smaller photometric H$\alpha$ lags and higher correlation coefficients. However, these trends are weak, indicating that the ICCF-Cut method is not very sensitive to these parameters. These simulations suggest that, for the cases considered here, the ICCF-Cut method can still recover broadly consistent lag measurements even when the continuum lag is neglected and the H$\alpha$ ratio is estimated from a single-epoch spectrum obtained during the monitoring period.
    
    \item For\,NGC\,4151\,(2022–2023),\,NGC\,4151\,(2024–2025), and\,UGC\,3374\,(2024–2025),\,the JAVELIN method produces lag distributions with multiple peaks or discrepant lags compared to the ICCF-Cut and $\chi^2$ methods. Through a WWZ analysis, we identify potential quasi-periodic signals in these light curves. Although these signals are not highly significant in all cases, their characteristic timescales appear broadly similar to the additional peaks in the JAVELIN lag distributions. The WWZ results therefore suggest a possible explanation for the multi-peaked lag distributions obtained with JAVELIN in these cases.

    \item For NGC 4151\,(2022–2023), UGC 3374\,(2023–2024), and UGC 3374 (2024–2025), we find that the extracted H$\alpha$ light curves are contaminated by the He\,\textsc{i} emission line. This contamination can bias the photometric H$\alpha$ lags toward the spectroscopic He\,\textsc{i} lags. Although the uncorrected ICCF-Cut lags are already broadly consistent with the spectroscopic results, we introduce and apply a refinement to explicitly subtract the He\,\textsc{i} component. This correction leads to a significant improvement in the agreement between the photometric and spectroscopic H$\alpha$ lags. These results suggest that contamination from the blended emission lines can influence broadband PRM measurements, and highly accurate lag measurements with the ICCF-Cut method may require careful source-specific contamination assessments and suitable spectroscopic priors.

    \item To further explore the applicability of the ICCF-Cut method under different observational conditions, we perform a PRM analysis using simulated light curves constructed with a variety of filters for NGC 4151 (2022–2023). For the simulated cases analyzed here, the ICCF-Cut method effectively mitigates continuum contamination and recovers H$\alpha$ lags broadly consistent with the spectroscopic measurements. In narrow-band PRM, the measured H$\alpha$ lags show the closest agreement with the spectroscopic results, whereas in broadband PRM, the improvement after continuum subtraction is the most pronounced. Overall, these simulations suggest that the ICCF-Cut method can perform well across different filter bandwidths explored in this work.
    
\end{enumerate}

\section*{ACKNOWLEDGEMENTS}
We are thankful for the support of the National Key R \& D Program of China (2025YFA1614100) and the National Natural Science Foundation of China (12133001, 12573021, 12303022, 12373018). We acknowledge the support of the staff of the Lijiang 2.4 m telescope. Funding for the telescope has been provided by Chinese Academy of Sciences and the People’s Government of Yunnan Province. We acknowledge the WFST team for providing the transmission functions of the photometric filters. This research has made use of the SVO Filter Profile Service "Carlos Rodrigo", funded by MCIN/AEI/10.13039/501100011033/ through grant PID2023-146210NB-I00.

\bibliography{sample701}{}
\bibliographystyle{aasjournalv7}

\appendix
\vspace{-20pt}

In this Appendix, we present the supplementary figures referred to in the main text. In Figure \ref{fig:NGC4151_UGC3374_LJT_result}, we present the photometric H$\alpha$ lag measurements for NGC 4151 and UGC 3374 in the remaining observing seasons (not shown in Figure \ref{fig:NGC4151_2223_result}). Figure \ref{fig:iccf_interband} and Figure \ref{fig:iccf_spec_results} display the detailed cross-correlation results for the inter-band lags and the reference spectroscopic lags, respectively, across all observed and simulated datasets. Furthermore, a direct comparison between the extracted ICCF-Cut photometric H$\alpha$ light curves and the corresponding spectroscopic H$\alpha$ light curves, along with their compared lags, is shown in Figure \ref{fig:Ha_Compare_All}. The photometric H$\alpha$ lag measurements based on the remaining simulated light curves (not shown in Figure \ref{fig:NGC4151_NAOC_C2_result}) of NGC 4151 (2022-2023) are shown in Figure \ref{fig:NGC4151_simulated_result}.

\renewcommand{\thefigure}{A\arabic{figure}}
\setcounter{figure}{0}

\begin{figure*}[h]
    \centering
    \begin{minipage}{0.5\textwidth}
        \includegraphics[width=\linewidth]{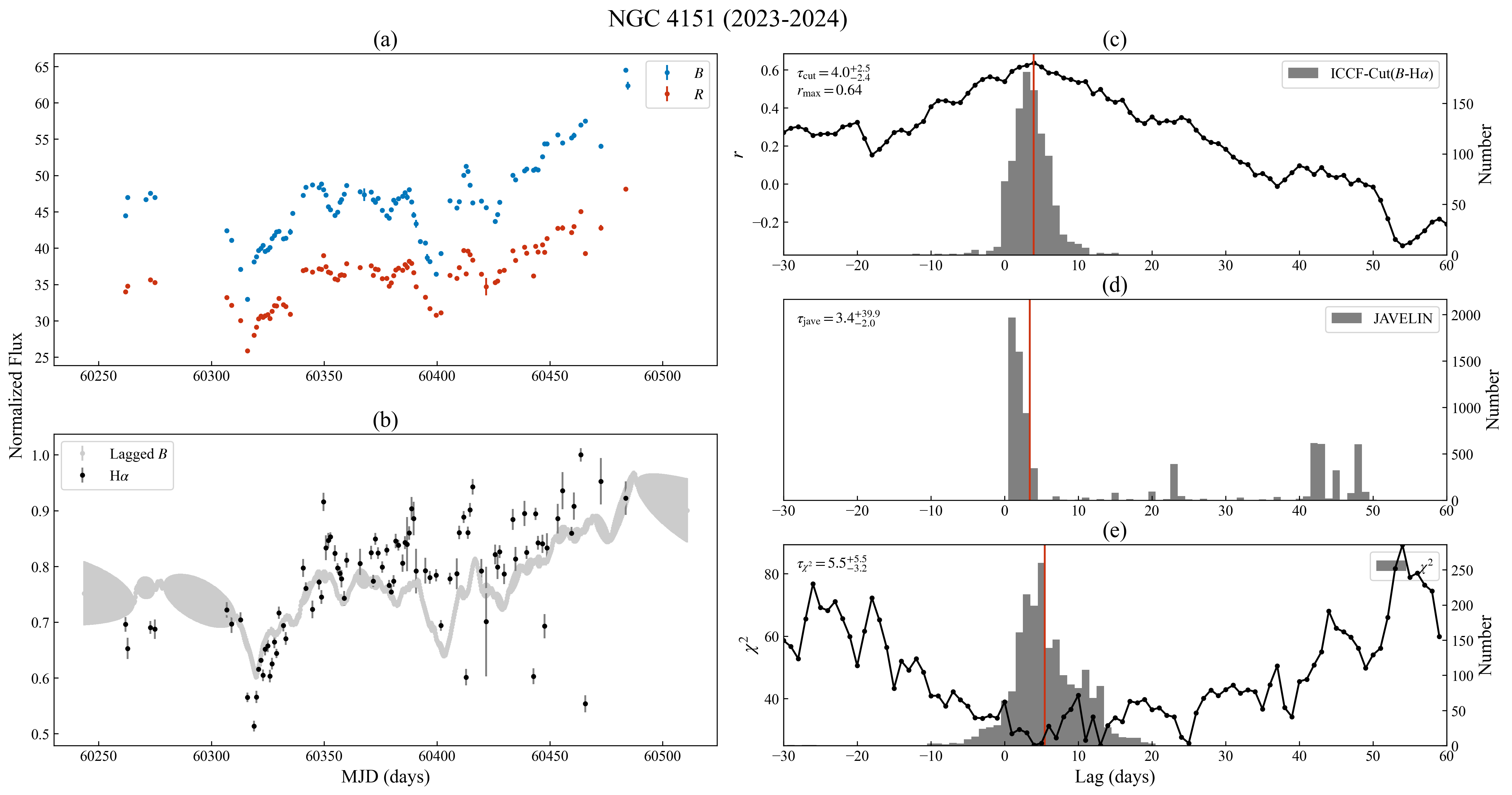}
    \end{minipage}%
    \begin{minipage}{0.5\textwidth}
        \includegraphics[width=\linewidth]{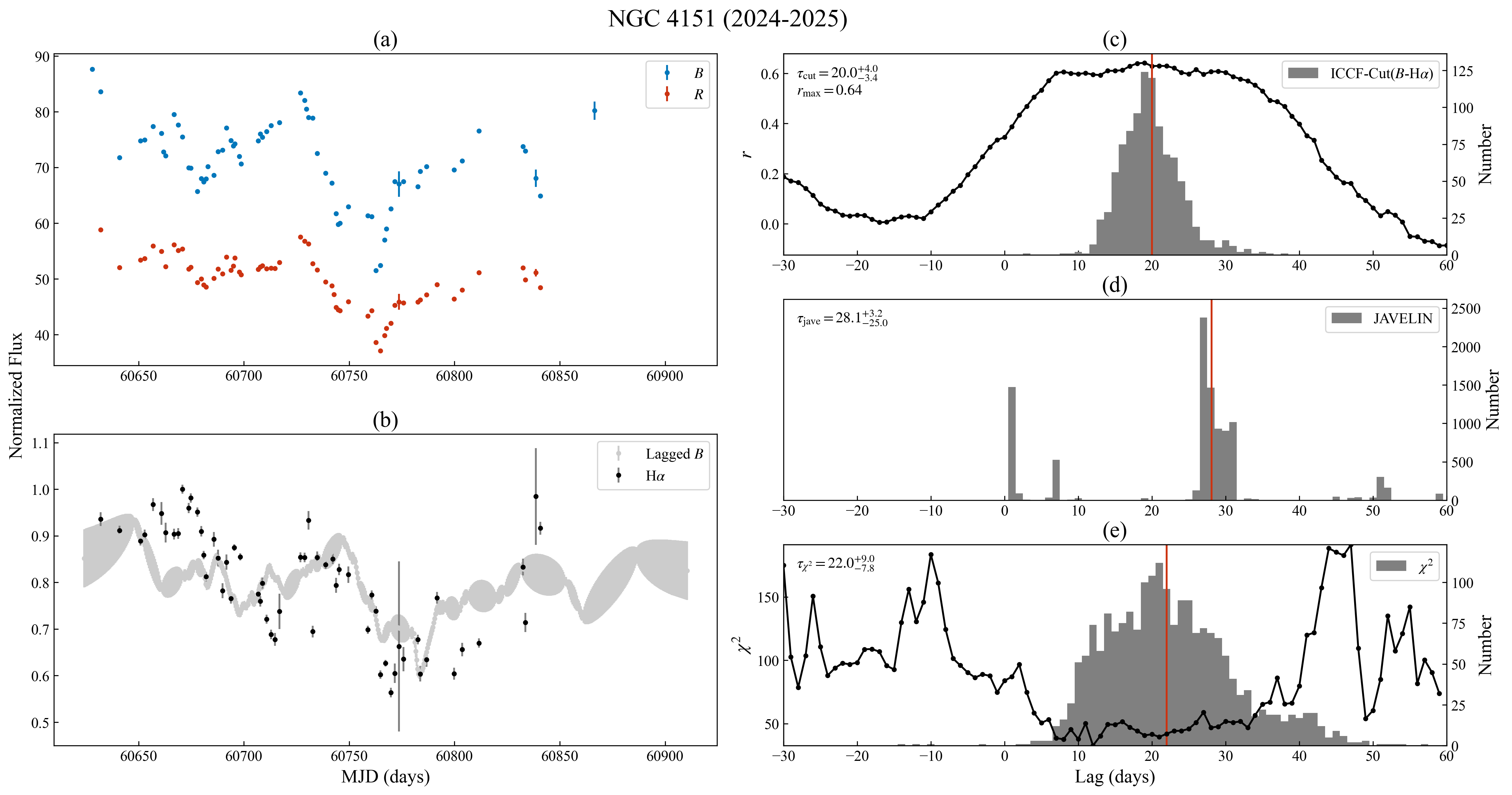}
    \end{minipage}

    \begin{minipage}{0.5\textwidth}
        \includegraphics[width=\linewidth]{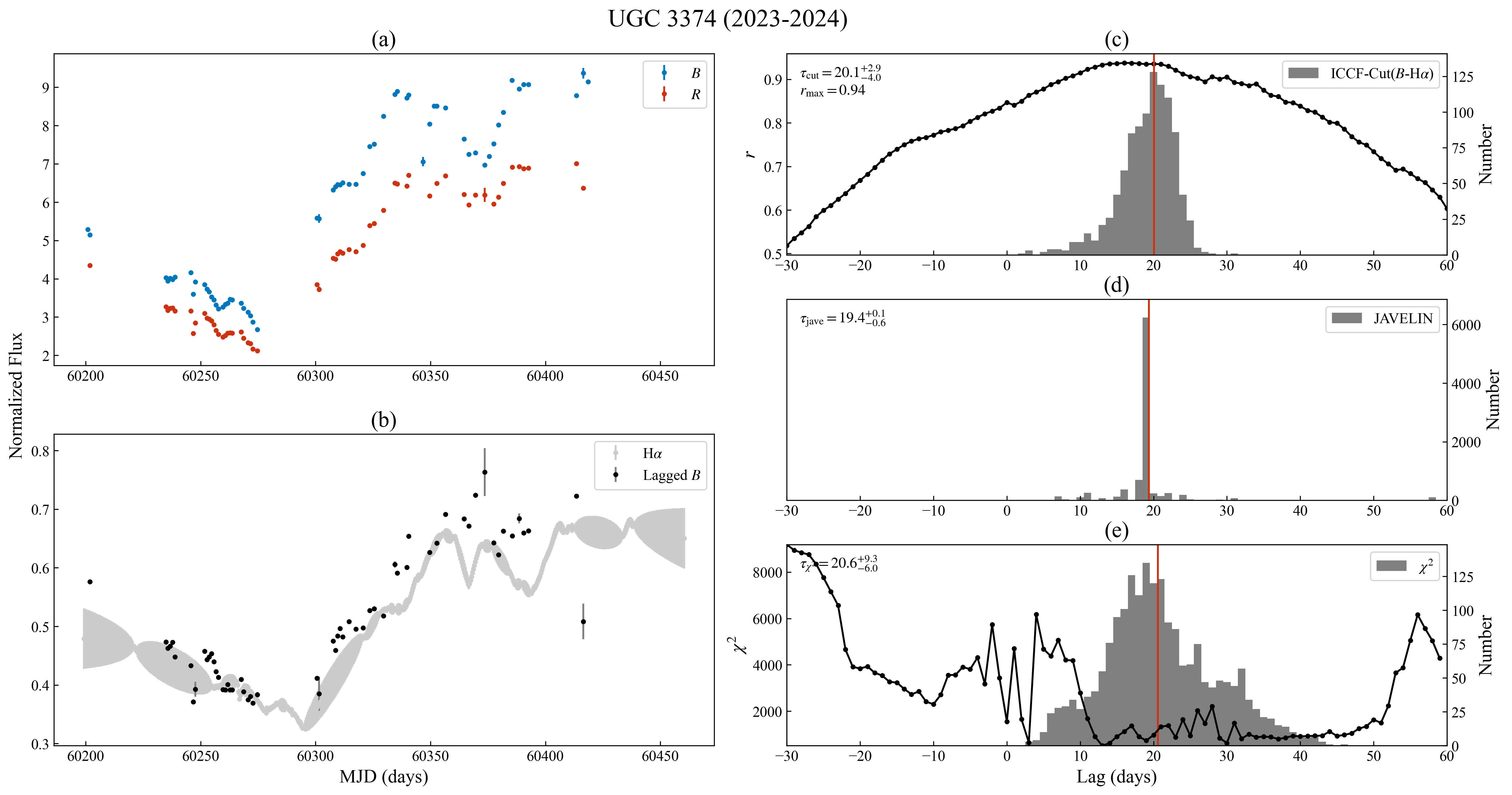}
    \end{minipage}%
    \begin{minipage}{0.5\textwidth}
        \includegraphics[width=\linewidth]{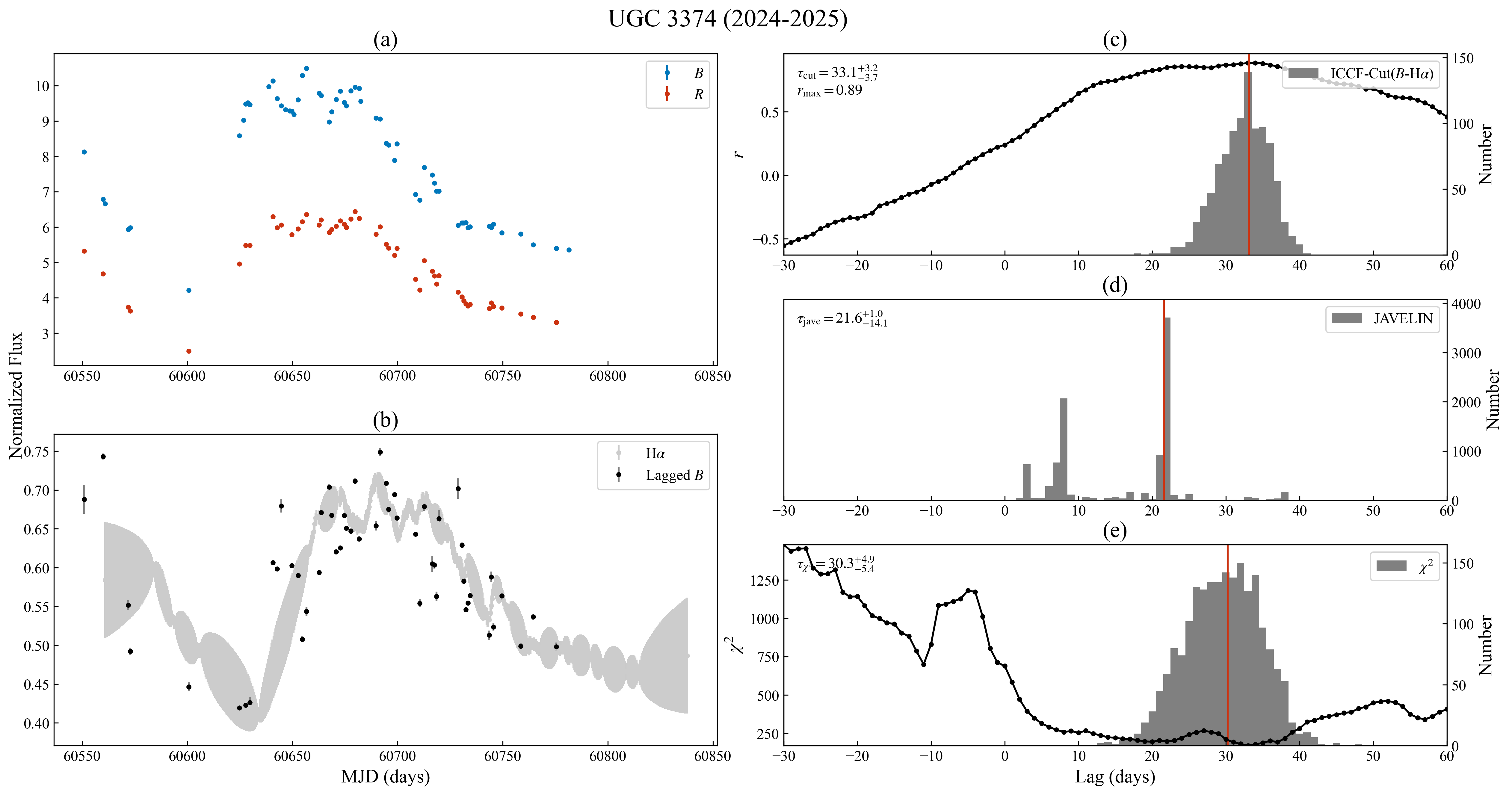}
    \end{minipage}
    \caption{Same as Figure \ref{fig:NGC4151_2223_result} but for NGC 4151 (LJT 2023-2024), NGC 4151 (LJT 2024-2025), UGC 3374 (LJT 2023-2024), and UGC 3374 (LJT 2024-2025).}
    \label{fig:NGC4151_UGC3374_LJT_result}
\end{figure*}

\begin{figure*}[h]
    \centering
    \includegraphics[width=1\textwidth]{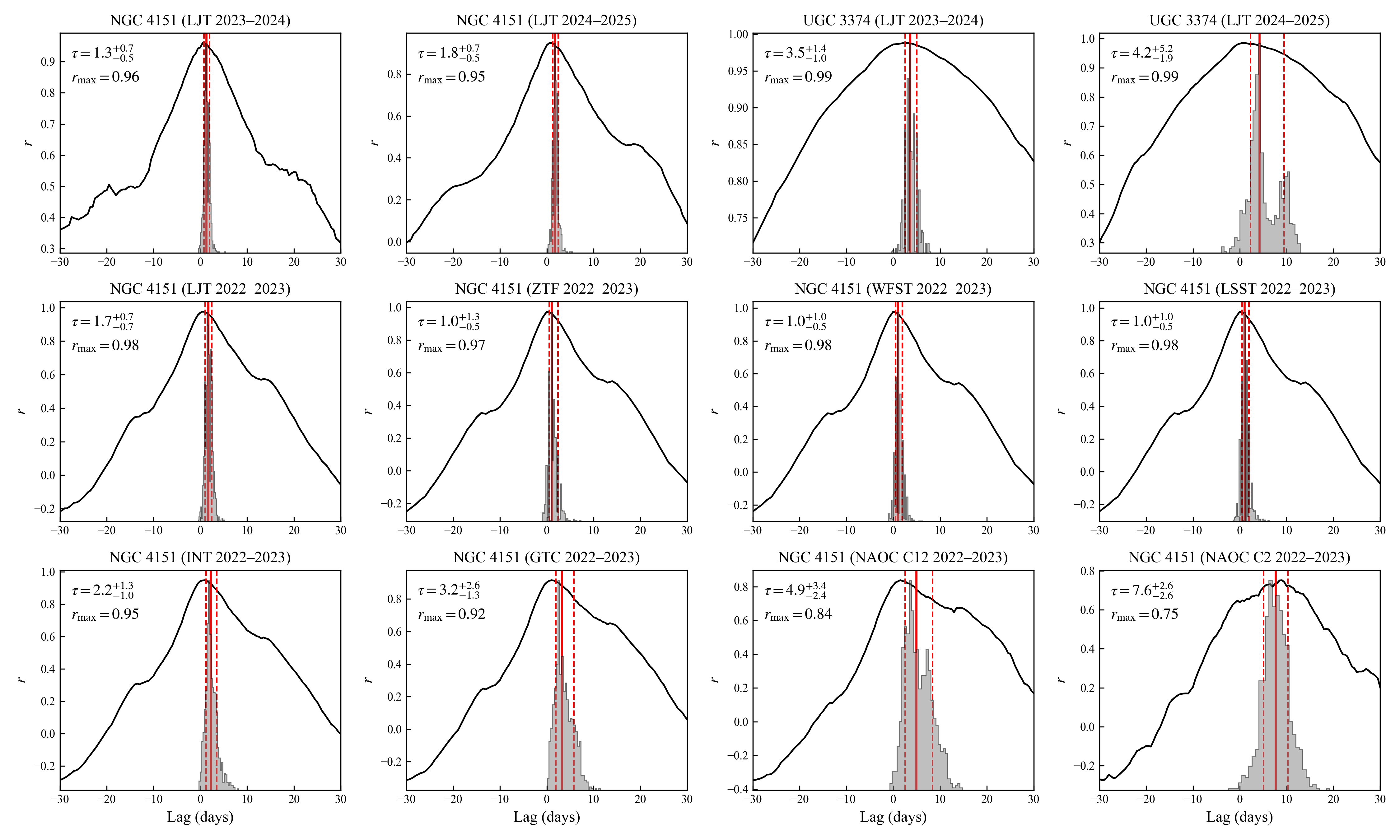} 
    \caption{Inter-band lag measurements between the continuum-band and line-band light curves. Each panel shows the CCF (black solid line) and the corresponding CCCD (gray shaded region). The red solid and dashed lines indicate the centroid lag and its $\pm1\sigma$ uncertainties, respectively. Top panels: Results for NGC 4151 (2023–2024, 2024–2025) and UGC 3374 (2023–2024, 2024–2025) using the observed $B$-band and $R$-band light curves from the LJT. Middle panels: Results for NGC 4151 (2022–2023) using the observed $B$-band and $R$-band light curves from the LJT, together with the simulated $g$-band and $r$-band light curves in the LJT, ZTF, WFST, and LSST filter systems. Bottom panels: Results for NGC 4151 (2022–2023) using the simulated light curves in the INT MR661, GTC F657, NAOC H$\alpha$ C12, and C2 filter systems, where the continuum band is represented by the observed LJT $B$-band light curve.
}
    \label{fig:iccf_interband}
\end{figure*}

\begin{figure*}[h]
    \centering
    \includegraphics[width=1\textwidth]{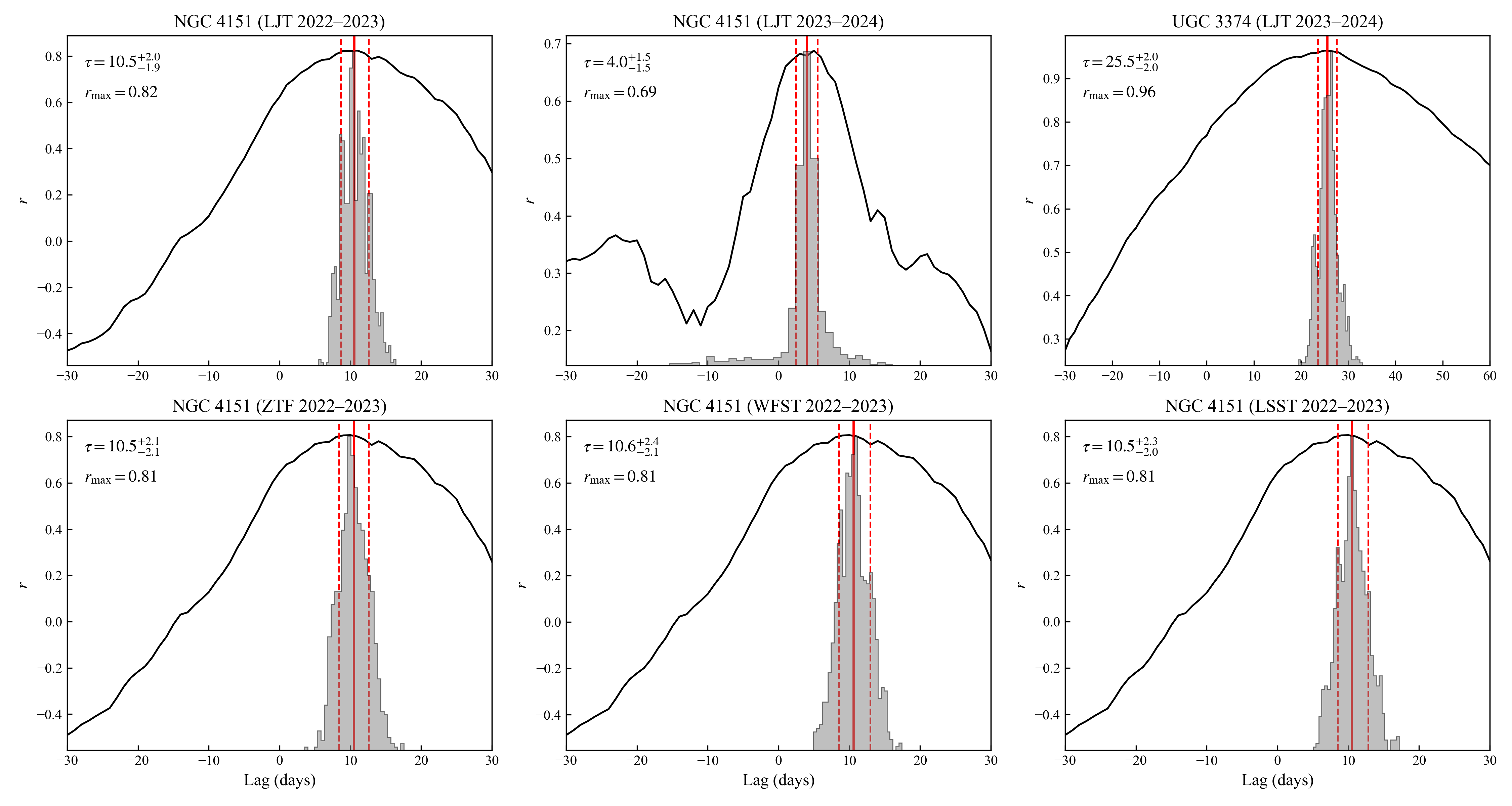} 
    \caption{Spectroscopic lag measurements between the photometric continuum-band and spectroscopic H$\alpha$ light curves. Each panel shows the ICCF (black solid line) and the corresponding CCCD (gray shaded region). The red solid and dashed lines indicate the centroid lag and its $\pm1\sigma$ uncertainties, respectively. Top panels: Results for NGC 4151 (2022–2023), NGC 4151 (2023–2024) and UGC 3374 (2023–2024) based on the LJT $B$-band photometric light curves and the LJT spectroscopic H$\alpha$ light curves. Bottom panels: Results for NGC 4151 (2022–2023) based on simulated $g$-band light curves corresponding to the ZTF, WFST, and LSST filter systems, cross-correlated with the LJT spectroscopic H$\alpha$ light curves.
    }
    \label{fig:iccf_spec_results}
\end{figure*}

\begin{figure*}[h]
    \centering
    \begin{minipage}[b]{0.46\textwidth}
        \includegraphics[width=\linewidth]{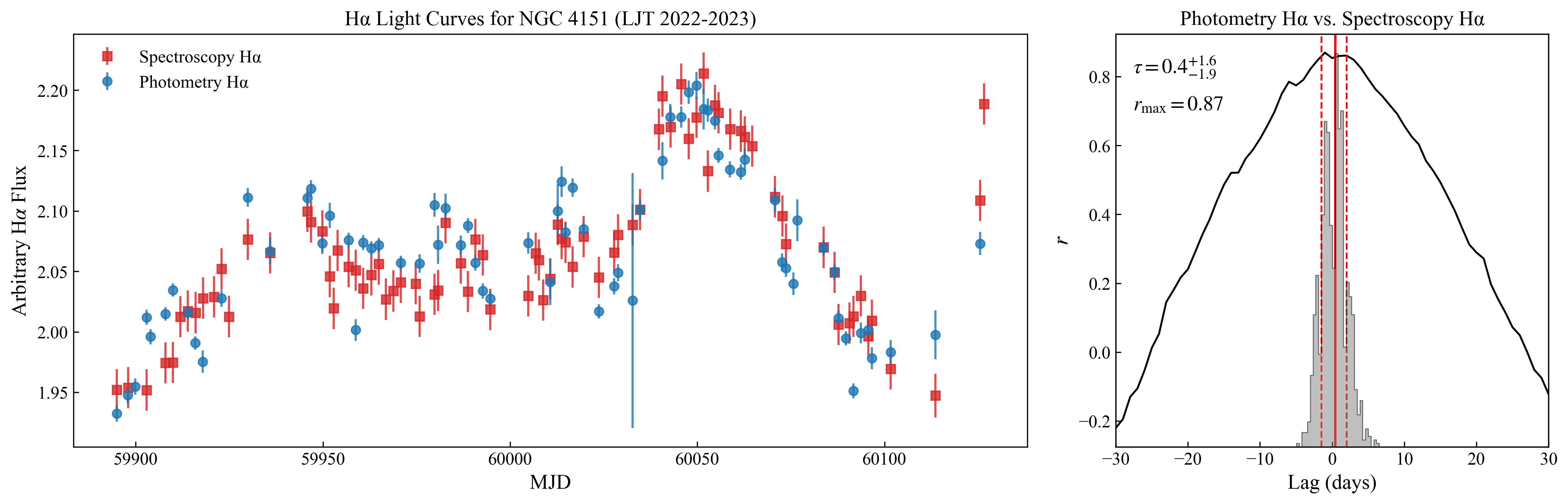}
    \end{minipage}\hfill
    \begin{minipage}[b]{0.46\textwidth}
        \includegraphics[width=\linewidth]{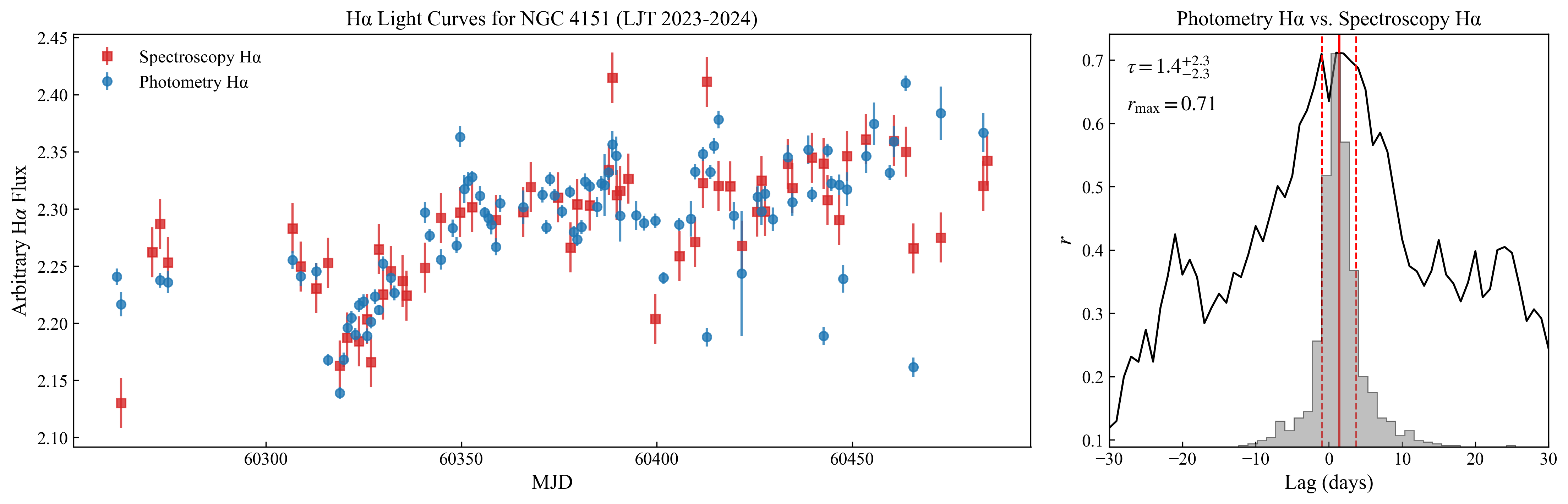}
    \end{minipage}

    \vspace{0mm}
    \begin{minipage}[b]{0.46\textwidth}
        \includegraphics[width=\linewidth]{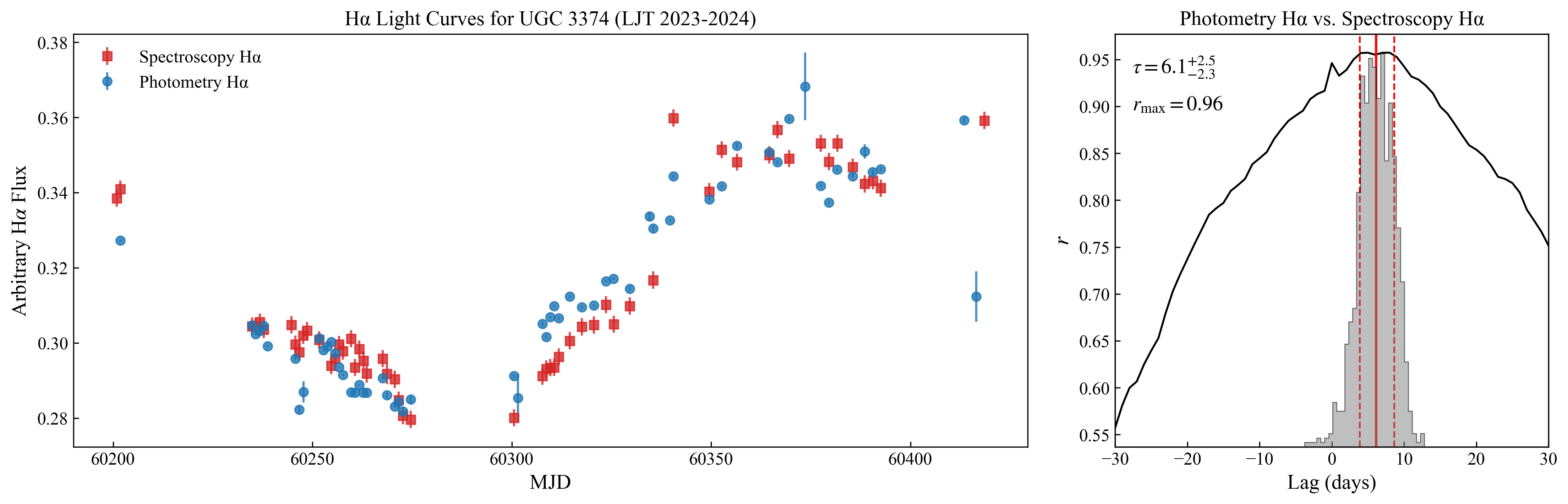}
    \end{minipage}\hfill
    \begin{minipage}[b]{0.46\textwidth}
        \includegraphics[width=\linewidth]{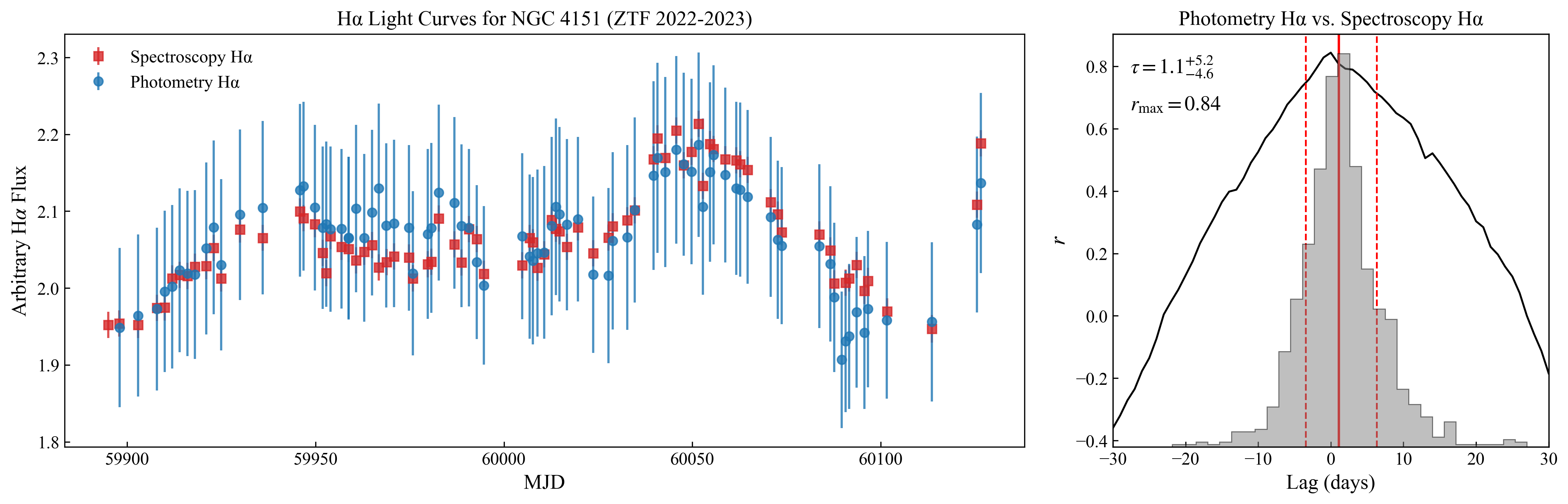}
    \end{minipage}

    \vspace{0mm}
    \begin{minipage}[b]{0.46\textwidth}
        \includegraphics[width=\linewidth]{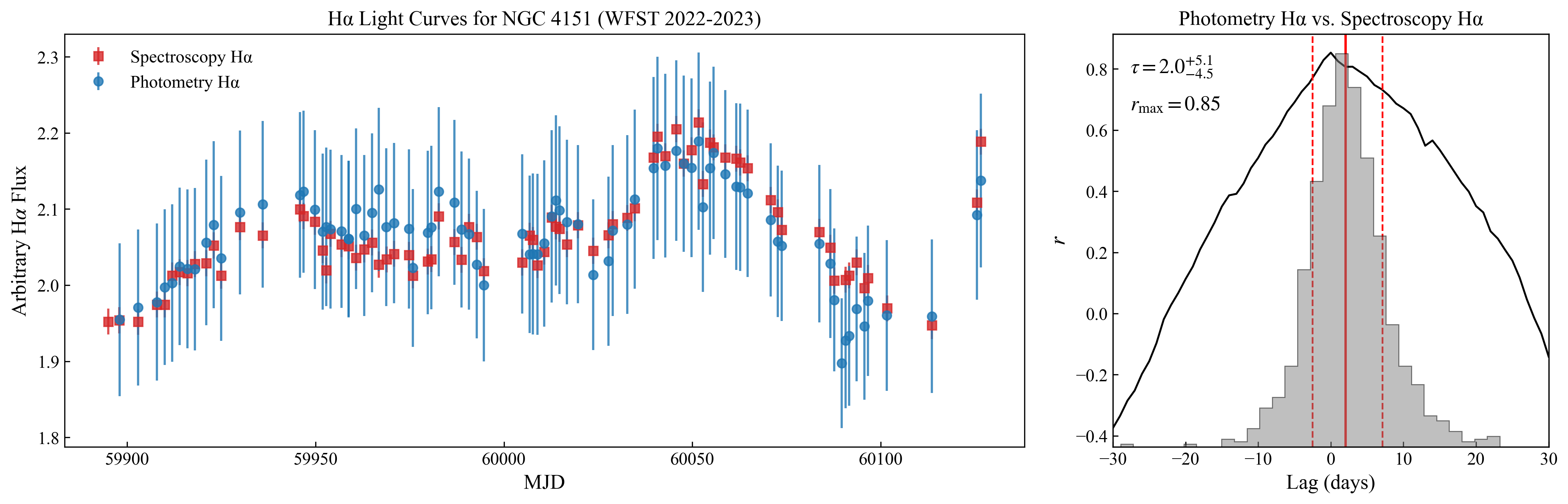}
    \end{minipage}\hfill
    \begin{minipage}[b]{0.46\textwidth}
        \includegraphics[width=\linewidth]{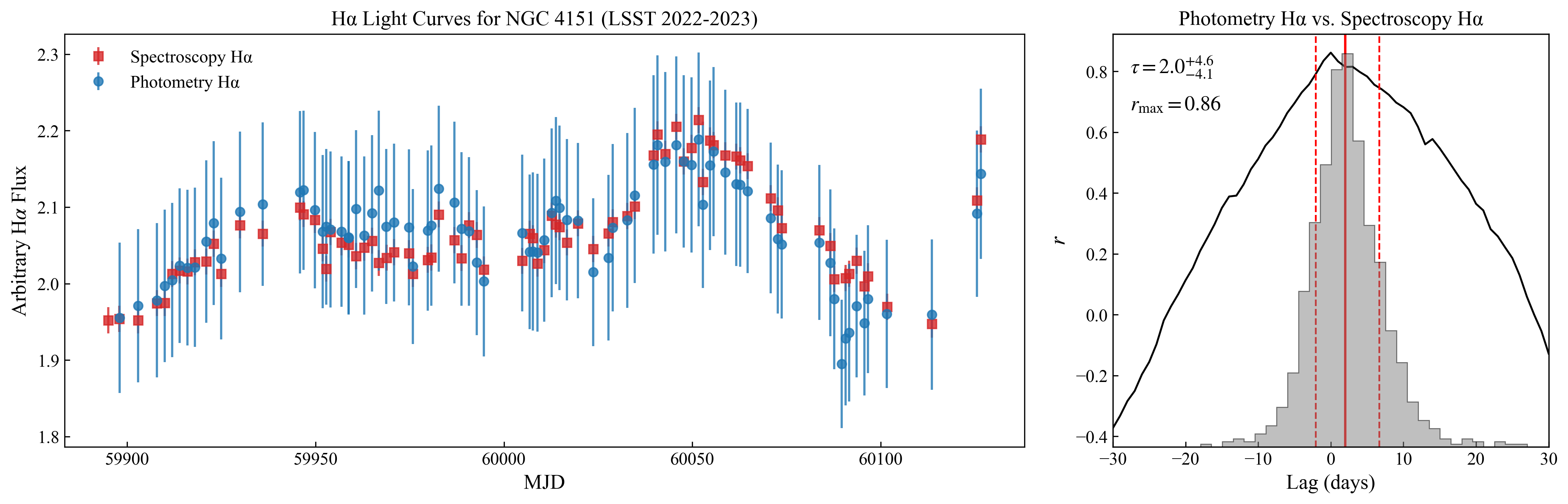}
    \end{minipage}

    \vspace{0mm}
    \begin{minipage}[b]{0.46\textwidth}
        \includegraphics[width=\linewidth]{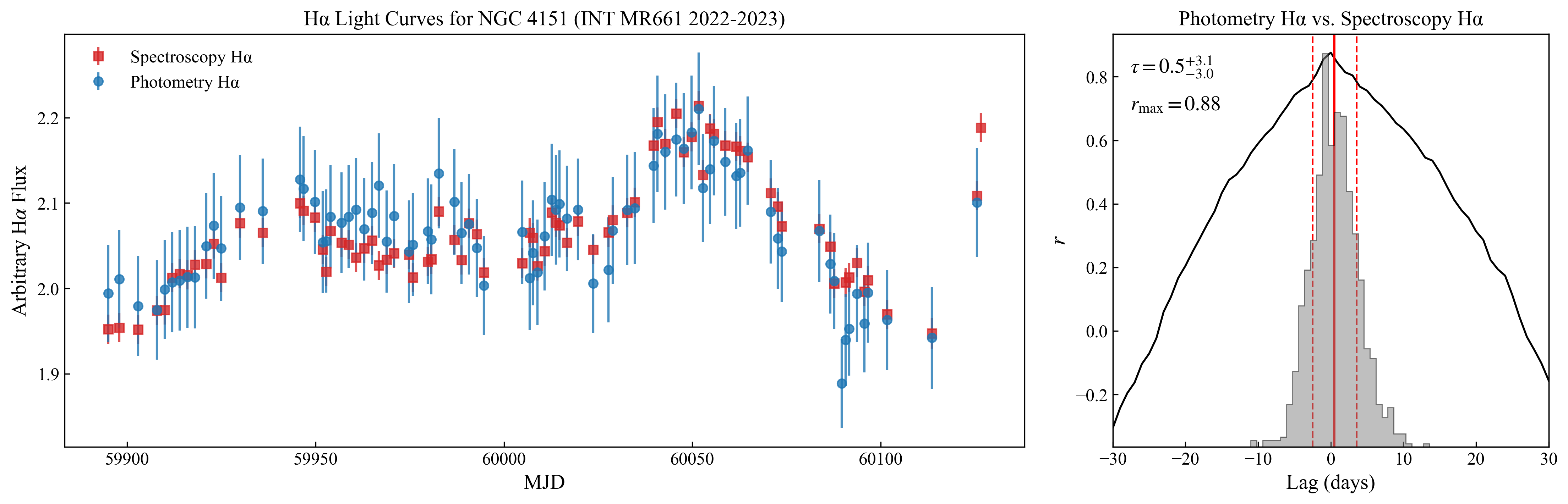}
    \end{minipage}\hfill
    \begin{minipage}[b]{0.46\textwidth}
        \includegraphics[width=\linewidth]{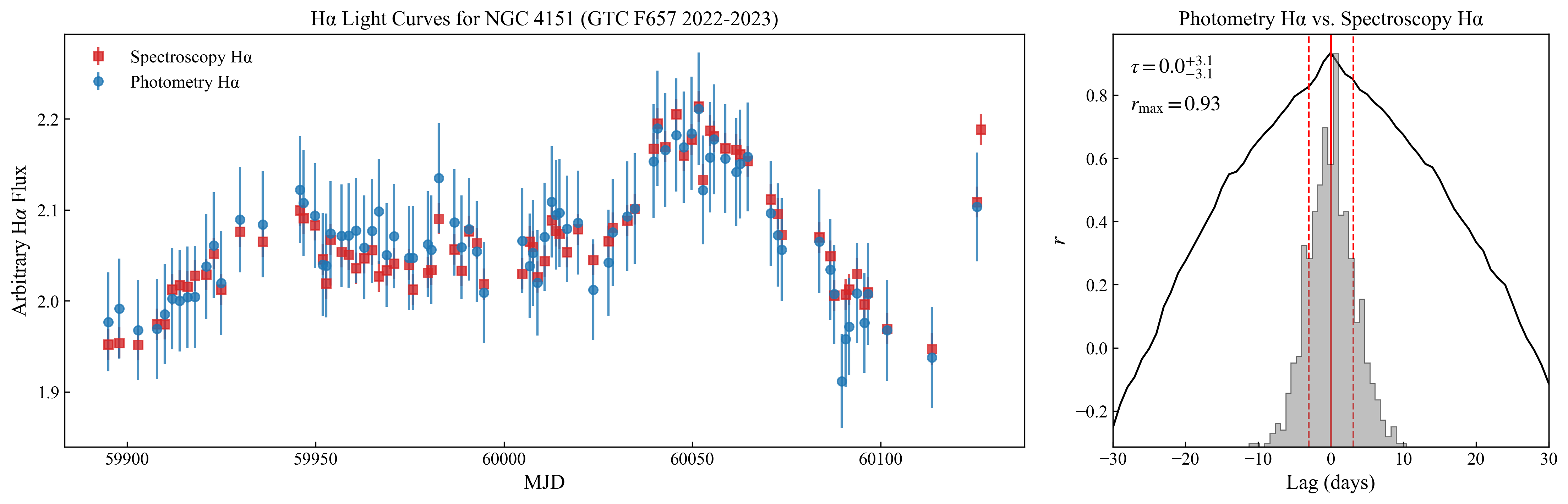}
    \end{minipage}

    \vspace{0mm}
    \begin{minipage}[b]{0.46\textwidth}
        \includegraphics[width=\linewidth]{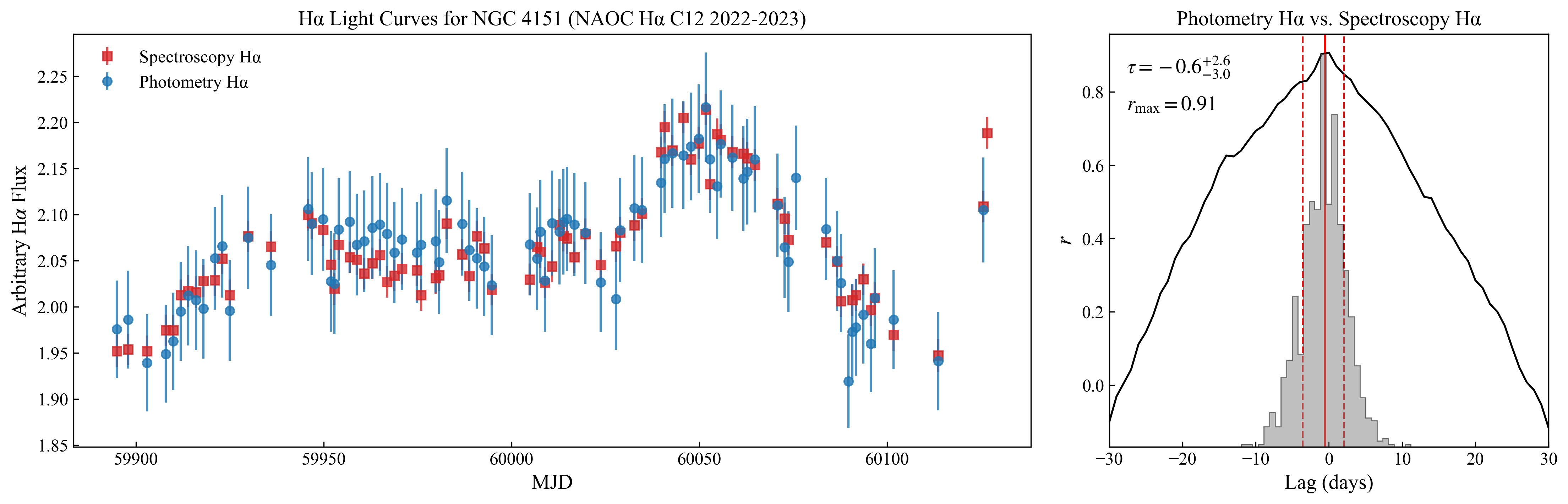}
    \end{minipage}\hfill
    \begin{minipage}[b]{0.46\textwidth}
        \includegraphics[width=\linewidth]{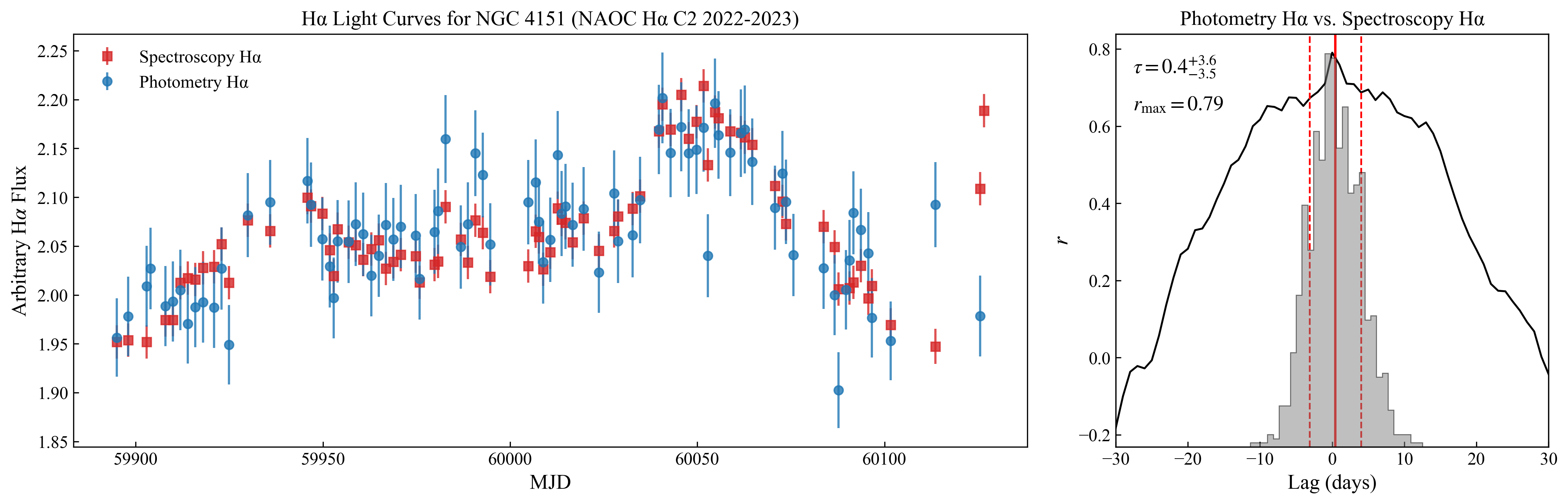}
    \end{minipage}

    \caption{
        Comparison between the ICCF-Cut photometric and spectroscopic H$\alpha$ light curves. The figure consists of ten subfigures (two columns by five rows). Each subfigure corresponds to a different photometric dataset for the ICCF-Cut analysis, with dataset names labeled above the left subpanels. In each subfigure, the left subpanel shows the ICCF-Cut photometric H$\alpha$ light curves (blue circles) and the spectroscopic H$\alpha$ light curves (red squares). The right subpanel displays the compared lag between them, with the ICCF (black solid line) and the corresponding CCCD (gray shaded region). Red solid and dashed lines indicate the centroid lag and its $\pm1\sigma$ uncertainties, respectively.
    }
    \label{fig:Ha_Compare_All}
\end{figure*}

\begin{figure*}[h]
    \centering
    \begin{minipage}{0.5\textwidth}
        \includegraphics[width=\linewidth]{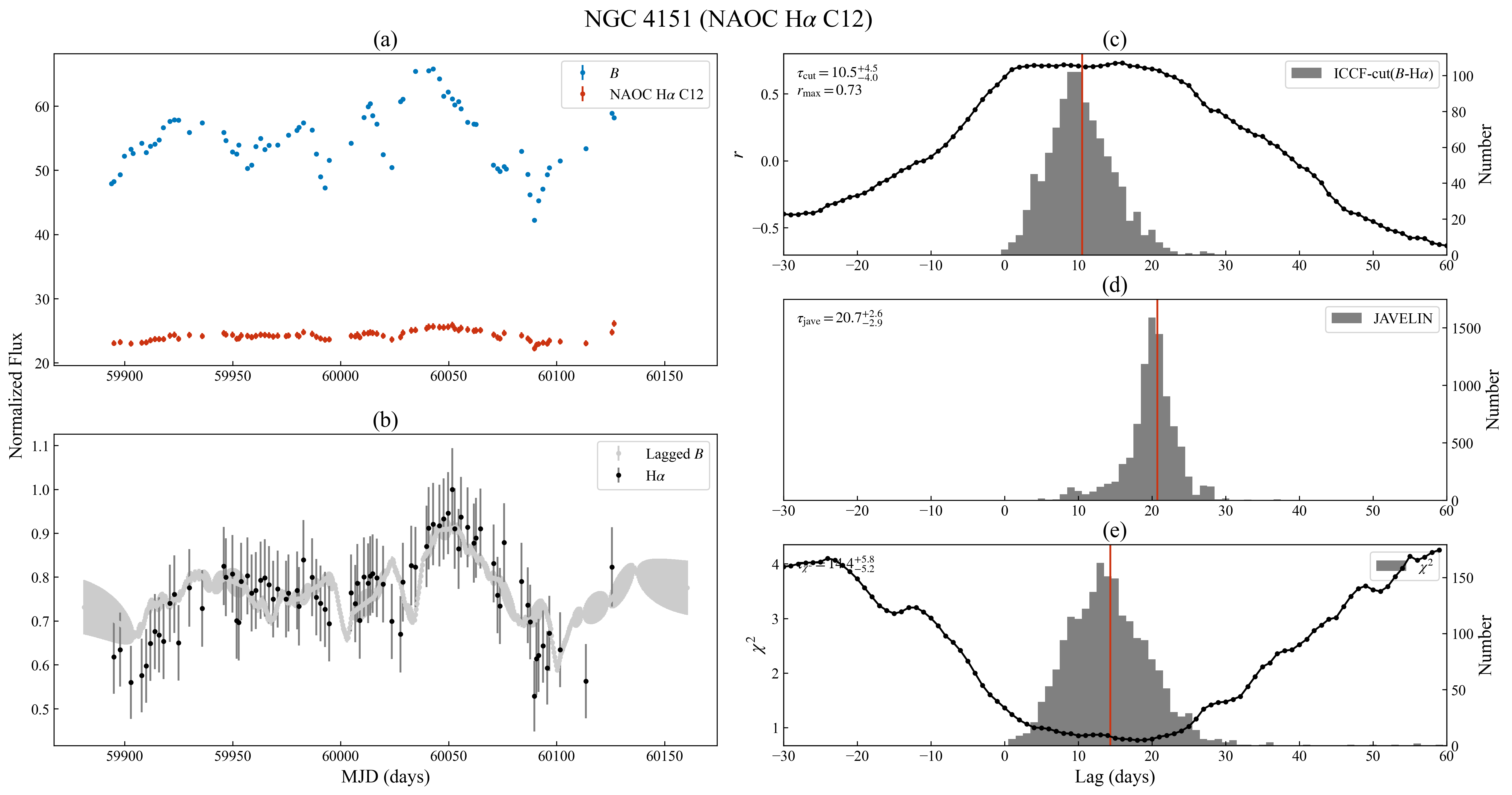}
    \end{minipage}%
    \begin{minipage}{0.5\textwidth}
        \includegraphics[width=\linewidth]{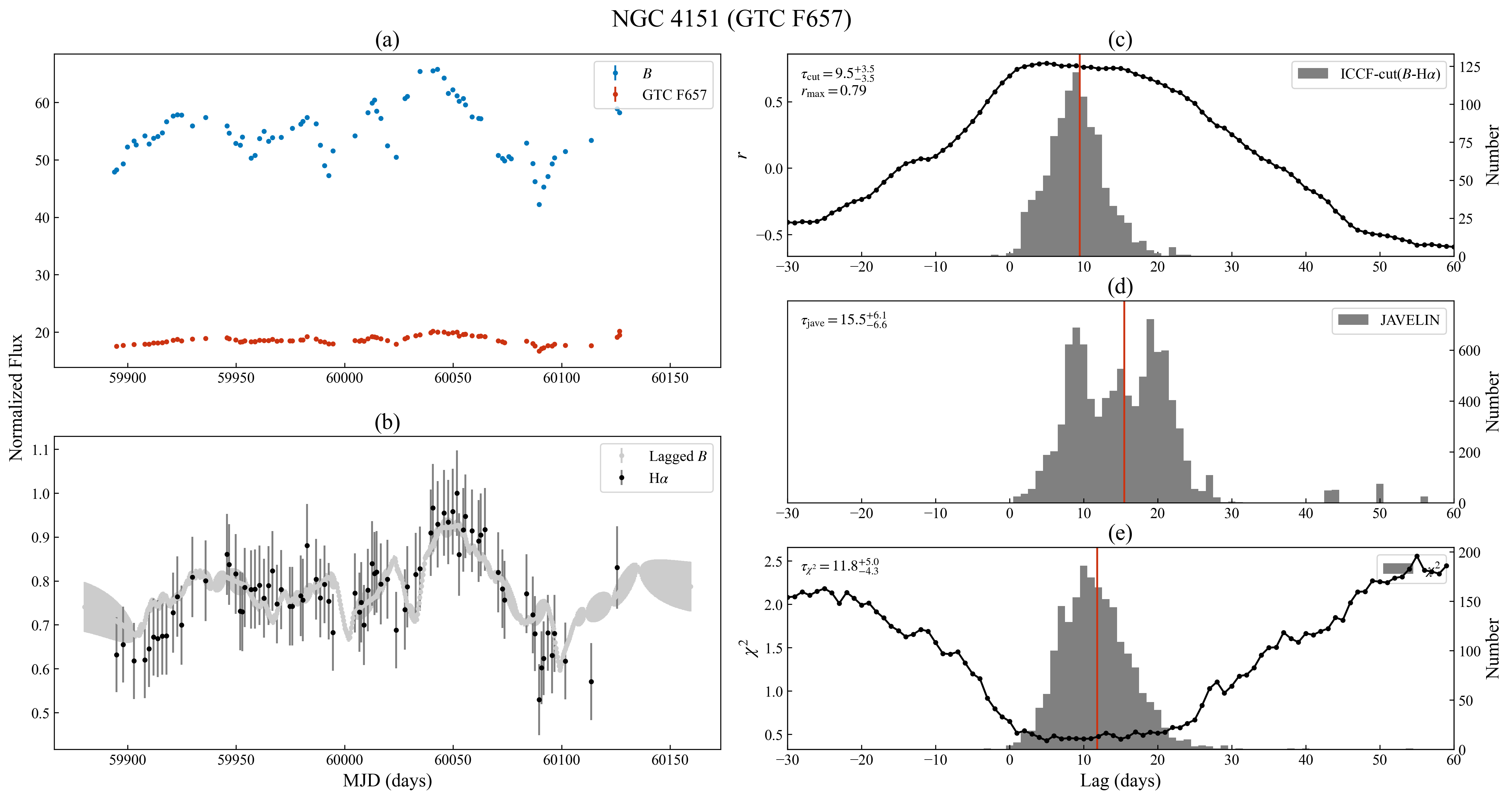}
    \end{minipage}

    \begin{minipage}{0.5\textwidth}
        \includegraphics[width=\linewidth]{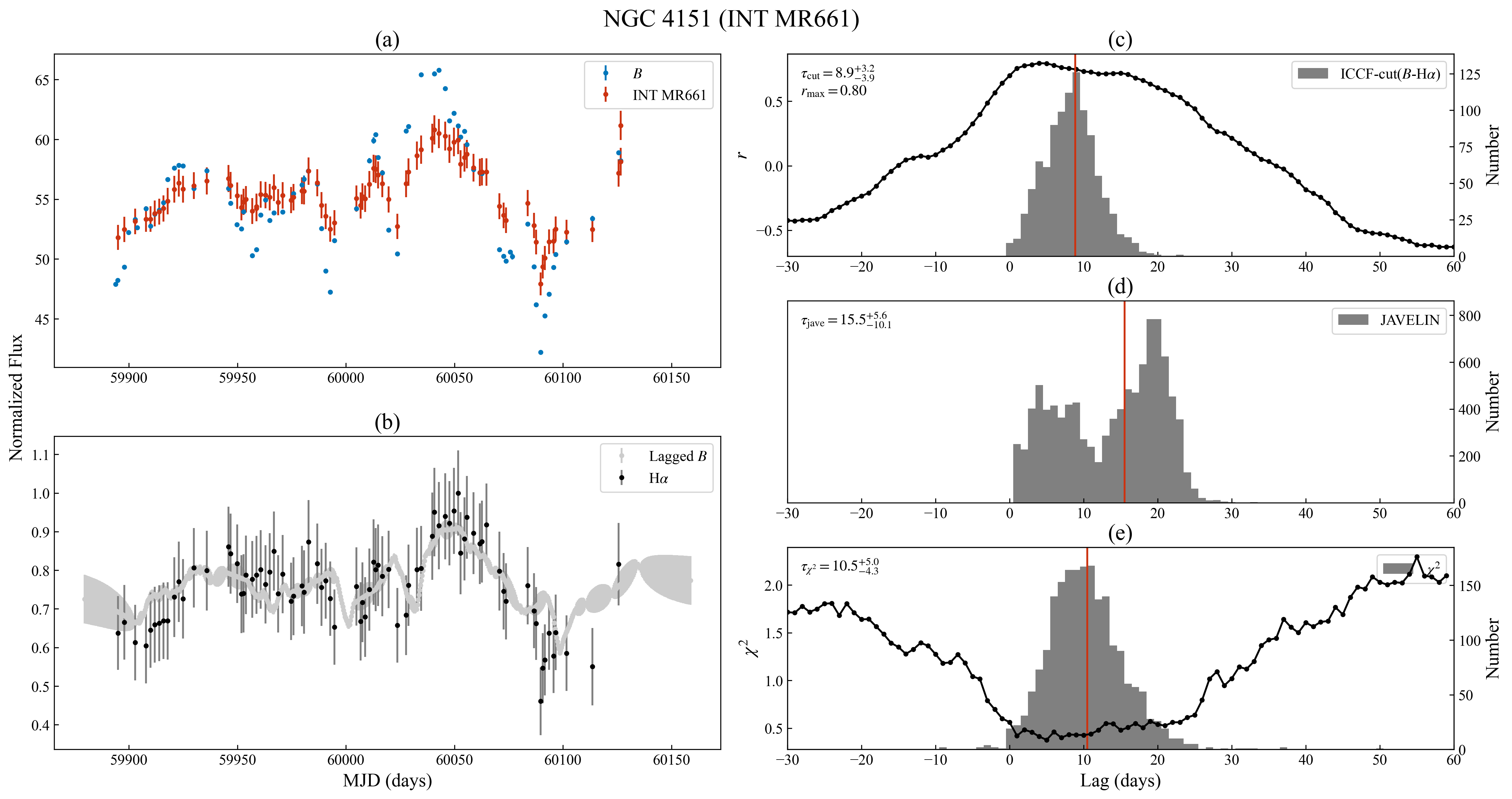}
    \end{minipage}%
    \begin{minipage}{0.5\textwidth}
        \includegraphics[width=\linewidth]{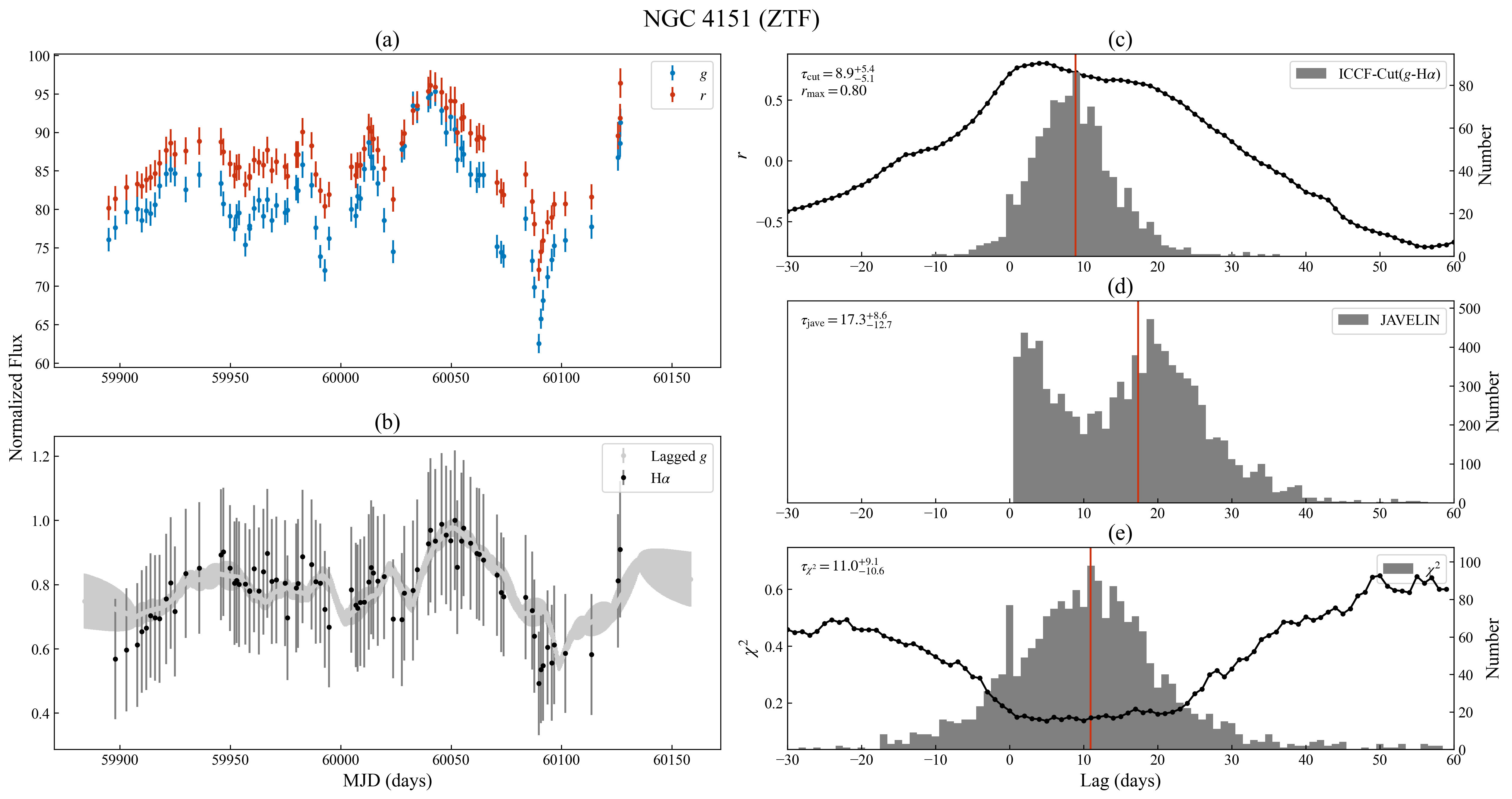}
    \end{minipage}

    \begin{minipage}{0.5\textwidth}
        \includegraphics[width=\linewidth]{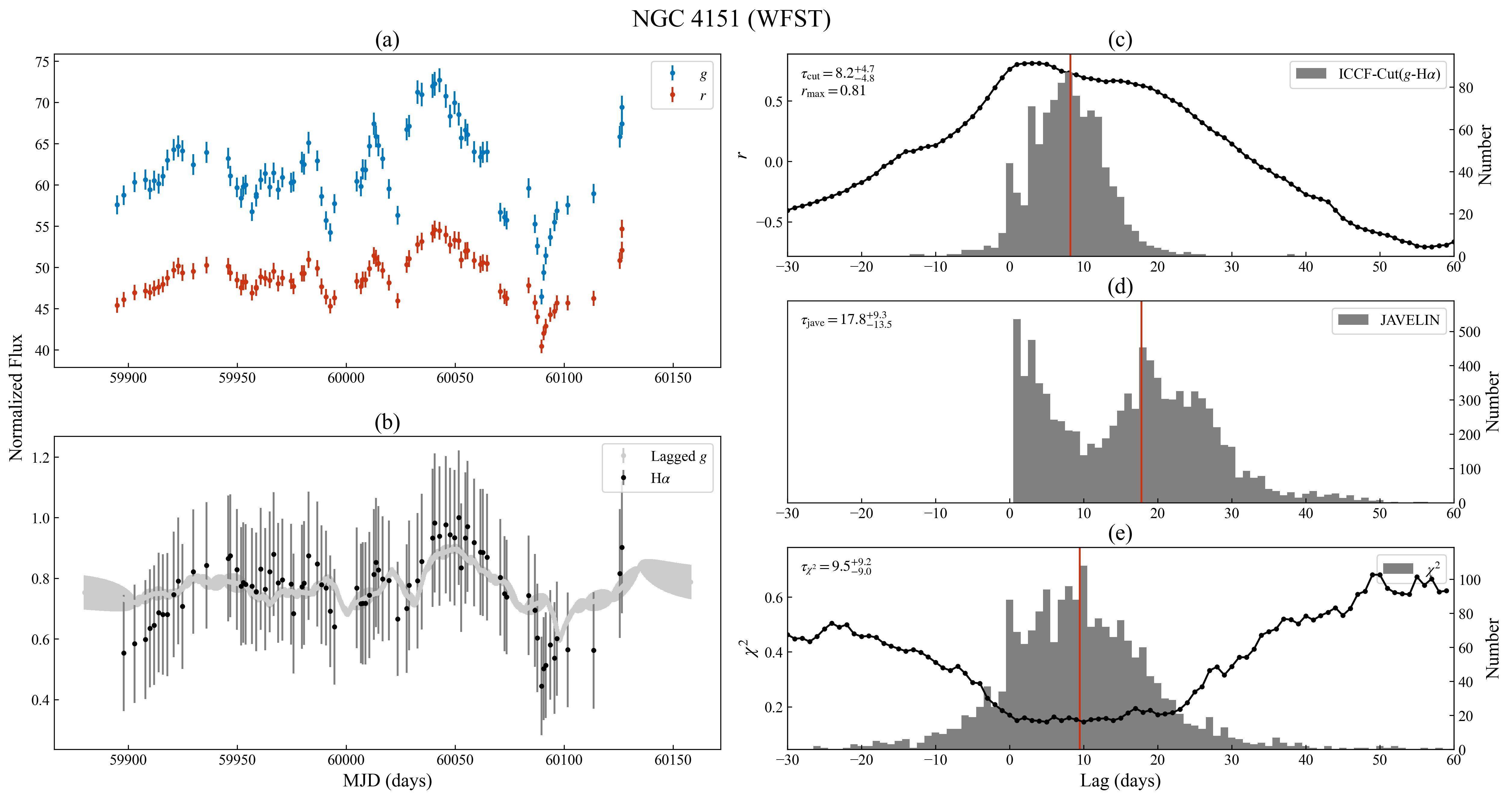}
    \end{minipage}%
    \begin{minipage}{0.5\textwidth}
        \includegraphics[width=\linewidth]{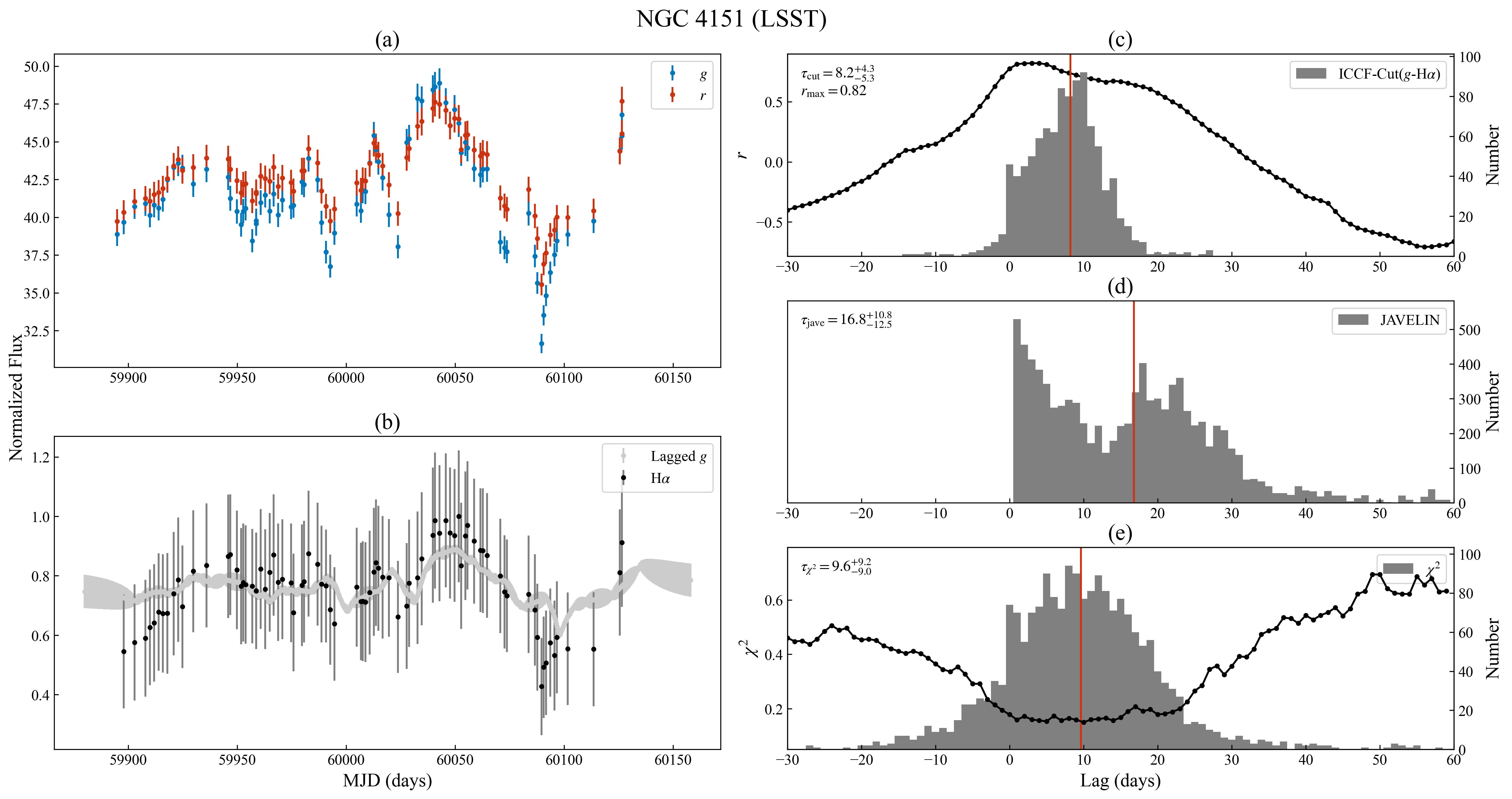}
    \end{minipage}
    \caption{Same as Figure \ref{fig:NGC4151_2223_result} but for NGC 4151 (NAOC H$\alpha$ C12 2022-2023), NGC 4151 (GTC 2022-2023), NGC 4151 (INT 2022-2023), NGC 4151 (ZTF 2022-2023), NGC 4151 (WFST 2022-2023), NGC 4151 (LSST 2022-2023).}
    \label{fig:NGC4151_simulated_result}
\end{figure*}

\end{document}